\documentclass[journal]{IEEEtran}
\ifCLASSINFOpdf
  \usepackage[pdftex]{graphicx}
  \graphicspath{{images/}}
  \DeclareGraphicsExtensions{.pdf,.jpeg,.png}
\else
\fi
\usepackage{bm}
\usepackage{svg}
\usepackage{amsmath}
\usepackage{amssymb}
\usepackage{algorithm}
\usepackage{algpseudocode}
\usepackage{placeins}
\usepackage{float}
\usepackage{hyperref}
\usepackage{soul}
\usepackage{booktabs}
\usepackage{diagbox}

\algnewcommand\algorithmicinput{\textbf{Input:}}
\algnewcommand\INPUT{\item[\algorithmicinput]}

\DeclareMathOperator{\EX}{\mathbb{E}} 

\newcommand{\bz}{\bm{z}}
\newcommand{\bx}{\bm{x}}
\newcommand{\by}{\bm{y}}
\newcommand{\bt}{\bm{\theta}}
\newcommand{\bp}{\bm{\phi}}
\newcommand{\bs}{\bm{\psi}}
\newcommand{\bxi}{\bm{x}_i}
\newcommand{\byi}{\bm{y}_i}

\begin{document}
%
\title{Adversarial Approximate Inference for Speech to Electroglottograph Conversion}
%
%
%

\author{Prathosh A. P.*, Varun Srivastava* and Mayank Mishra* \thanks{* Equal Contribution \newline All authors are with Indian Institute of Technology Delhi, New-Delhi - 110016, India. (e-mail: prathoshap@iitd.ac.in, varunsrivastava.v@gmail.com, mayank31398@gmail.com). The code used in this paper can be accessed via the following github link \url{https://github.com/VarunSrivastavaIITD/AAI}}}

%
%

\markboth{IEEE/ACM Transactions on Audio Speech and Language Processing}%
{Shell \MakeLowercase{\textit{et al.}}: Bare Demo of IEEEtran.cls for IEEE Journals}
%



\maketitle

\begin{abstract}

	Speech produced by human vocal apparatus conveys substantial non-semantic information including the gender of the speaker, voice quality, affective state, abnormalities in the vocal apparatus etc. Such information is attributed to the properties of the voice source signal, which is usually estimated from the speech signal. However, most of the source estimation techniques depend heavily on the goodness of the model assumptions and are prone to noise. A popular alternative is to indirectly obtain the source information through the Electroglottographic (EGG) signal that measures the electrical admittance around the vocal folds using dedicated hardware. In this paper, we address the problem of estimating the EGG signal directly from the speech signal, devoid of any hardware. Sampling from the intractable conditional distribution of the EGG signal given the speech signal is accomplished through optimization of an evidence lower bound. This is constructed via minimization of the KL-divergence between the true and the approximated posteriors of a latent variable learned using a deep neural auto-encoder that serves an informative prior. We demonstrate the efficacy of the method at generating the EGG signal by conducting several experiments on datasets comprising multiple speakers,  voice qualities, noise settings and speech pathologies. The proposed method is evaluated on many benchmark metrics and is found to agree with the gold standard while proving better  than  the  state-of-the-art  algorithms  on  a  few  tasks such  as  epoch  extraction.
\end{abstract}


%
\IEEEpeerreviewmaketitle

\section{Introduction}
%
%
%
%

\subsection{Background}

\IEEEPARstart{H}{uman} speech is often said to convey information on several levels \cite{laver1994principles}, broadly categorized as linguistic, para-linguistic and extra-linguistic layers.  The semantic and grammatical content of the intended text is encoded in the linguistic layer through the phonetic units. Significant non-lexical information including the speaker-dialect, emotional and affective state is embedded in the para-linguistic layer. The extra-linguistic layer encompasses the physical and physiological properties of the speaker and the vocal apparatus. This includes the identity, gender, age, prosody, loudness and other characteristics of the speaker and the physiological status of the vocal apparatus such as the manner of phonation, presence of vocal disorders etc. \cite{marasek1997egg}. The extra-linguistic information is sometimes also referred to as the Voice Quality (VQ), which is primarily characterized by laryngeal and supralaryngeal features arising during phonation \cite{trask2004dictionary}.
The perceived VQ largely depends upon the phonation, namely the process of converting a quasi-periodic respiratory air-flow into audible speech through the vibrations of vocal folds \cite{ladefoged1988investigating}.
Different types of laryngeal functions and configurations give rise to different phonation types. A few examples include breathy voice, falsetto, creaky and pathological voices that refer to the abnormalities in the vocal folds  \cite{hirano1981clinical}.

\begin{figure}[h!]
	\includegraphics[width=0.45\textwidth, height=0.3 \textwidth]{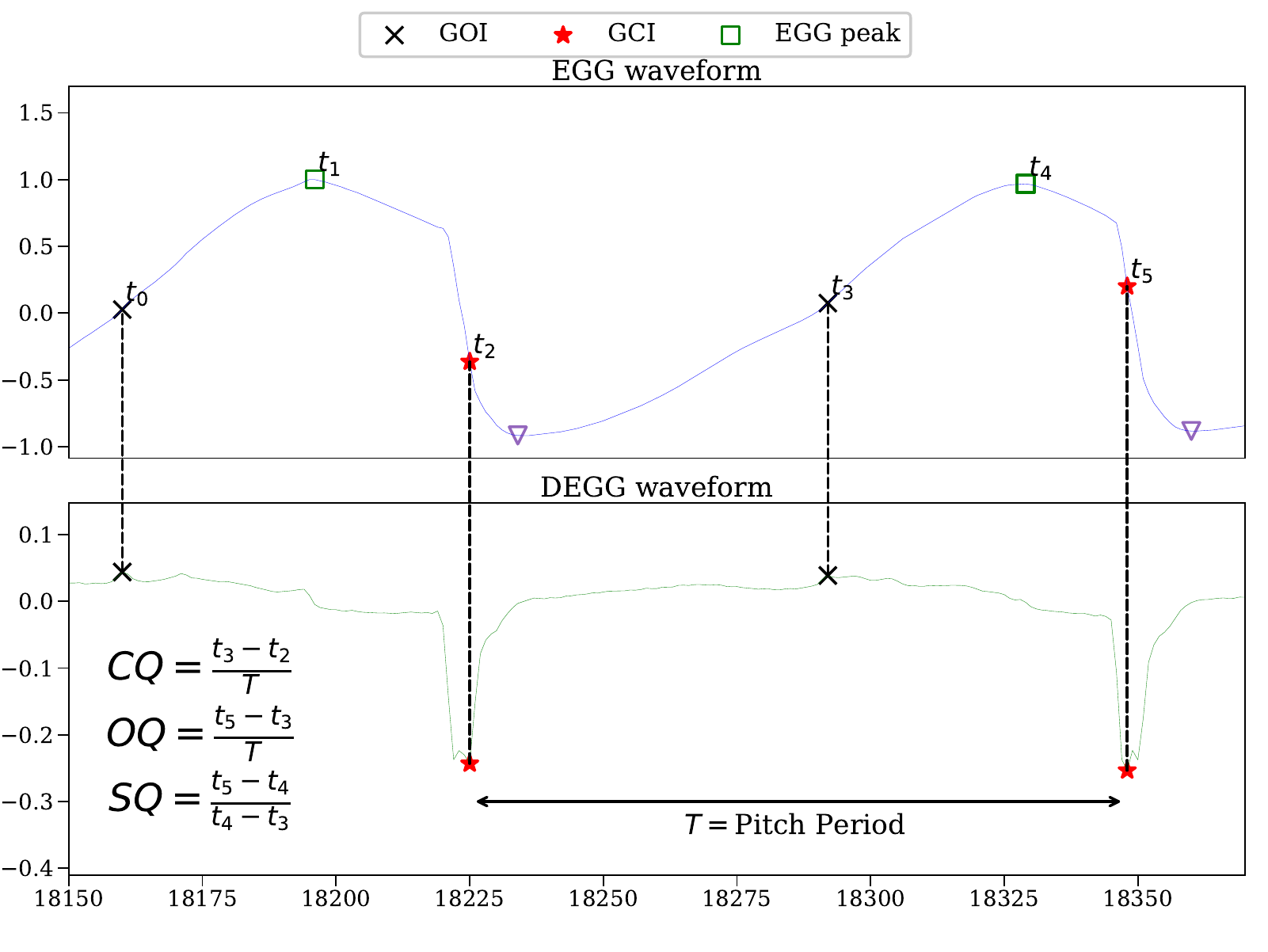}
	\centering
	\caption{Depiction of two EGG cycles and its derivative (DEGG) with the markings of the corresponding epochs.}
	\label{fig:egg_cycle_bdl}
\end{figure}

\subsection{Characterization of laryngeal behaviour}

Given its wide applicability in speech analysis, it is desirable to characterize laryngeal behaviour. A wide range of methods exist in practice to do so, starting from simple listening tests to invasive optical technique such as laryngeal stroboscopy \cite{kitzing1985stroboscopy}. One of the most popular class of approaches to characterize laryngeal behaviour is to use speech signal itself to get an estimate of the voice source signal.
\cite{gobl1992acoustic,wong1979least,wakita1973direct,airaksinen2014quasi,airaksinen2017quadratic,rao2019glottal,raitio2011hmm}.
This is usually accomplished using the inverse filtering technique (hence the name glottal inverse filtering) under the assumptions specified by the linear source-filter model for speech production \cite{makhoul1975linear}. This method is completely non-invasive and requires no additional sensing hardware since it estimates the glottal activity directly from the speech signal. However, these methods heavily depend on the correctness of the model assumptions, accurate estimation of formant frequencies and closed phase bandwidths. This is especially true for high pitched, nasalized and pathological voices. Further the estimated glottal source is known to suffer from the presence of ripples due to improper formant cancellation especially when there is noise in the recording environment or in the voice production mechanism itself \cite{alku2011glottal}. This leads to improper estimation of underlying voice quality measures (Refer Fig. 2 for a depiction this). A popular alternative is electroglottography which is a non-invasive technique to estimate the laryngeal behaviour devoid of the aforementioned limitations of the glottal inverse filtering \cite{veeneman1985automatic}.

\subsection{Laryngography - An introduction}

Laryngography or Electroglottography (EGG) is a technique used to indirectly quantify laryngeal behavior by  measuring the change in electrical impedance across the throat during speaking \cite{baken1992electroglottography,childers1985critical}. A low voltage high frequency current signal is fed through the larynx and the change in the impedance, which depends on the vocal fold contact properties, is recorded \cite{titze1990interpretation}. EGG is shown to have multiple correlates of laryngeal behaviour and has proven useful in multiple tasks such as assessment of phonation types \cite{peterson1994comparison,liu2017comparison}, gender \cite{chen2002electroglottographic,higgins2002gender} , emotional state and \cite{waaramaa2013acoustic,murphy2009electroglottogram}, identification of voice pathologies \cite{kitzing1985stroboscopy}.

The amplitude of the EGG waveform is known to vary linearly with the vocal fold contact area \cite{schered1987electroglottography}. Thus different sections of the EGG mark different laryngeal activities. Further, the EGG waveform is quasi-periodic during the production of voiced phonemes and has zero or very low amplitude during the unvoiced counterparts. The duration between successive positive peaks in the EGG signal correspond to the instantaneous pitch period and are called glottal cycles. For most of the aforementioned use cases, EGG is analyzed cycle-wise and hence parameterized in accordance with a few epochal points within each cycle. The significant quantification measures for the EGG are the instants of glottal opening, glottal closure, location of the maximum glottal opening and start and end of the cycles (Refer Fig. \ref{fig:egg_cycle_bdl} for a sample EGG waveform for two glottal cycles with the corresponding epochs marked). Further, duration of certain events relative to the pitch period such as glottal closure, opening and the skewness of the waveform in every cycle are known to signify several properties of the glottal activity \cite{liu2017comparison,deshpande2018effective}. These are respectively called the contact/closure, open, and speed quotients (Refer Fig. \ref{fig:egg_cycle_bdl}). Furthermore, the morphology of the entire EGG waveform over longitudinal cycles serves several clinical applications \cite{kitzing1990clinical}.

\begin{figure}[h!]
	\includegraphics[trim={6cm 1cm 4cm 0},clip,width=0.45\textwidth]{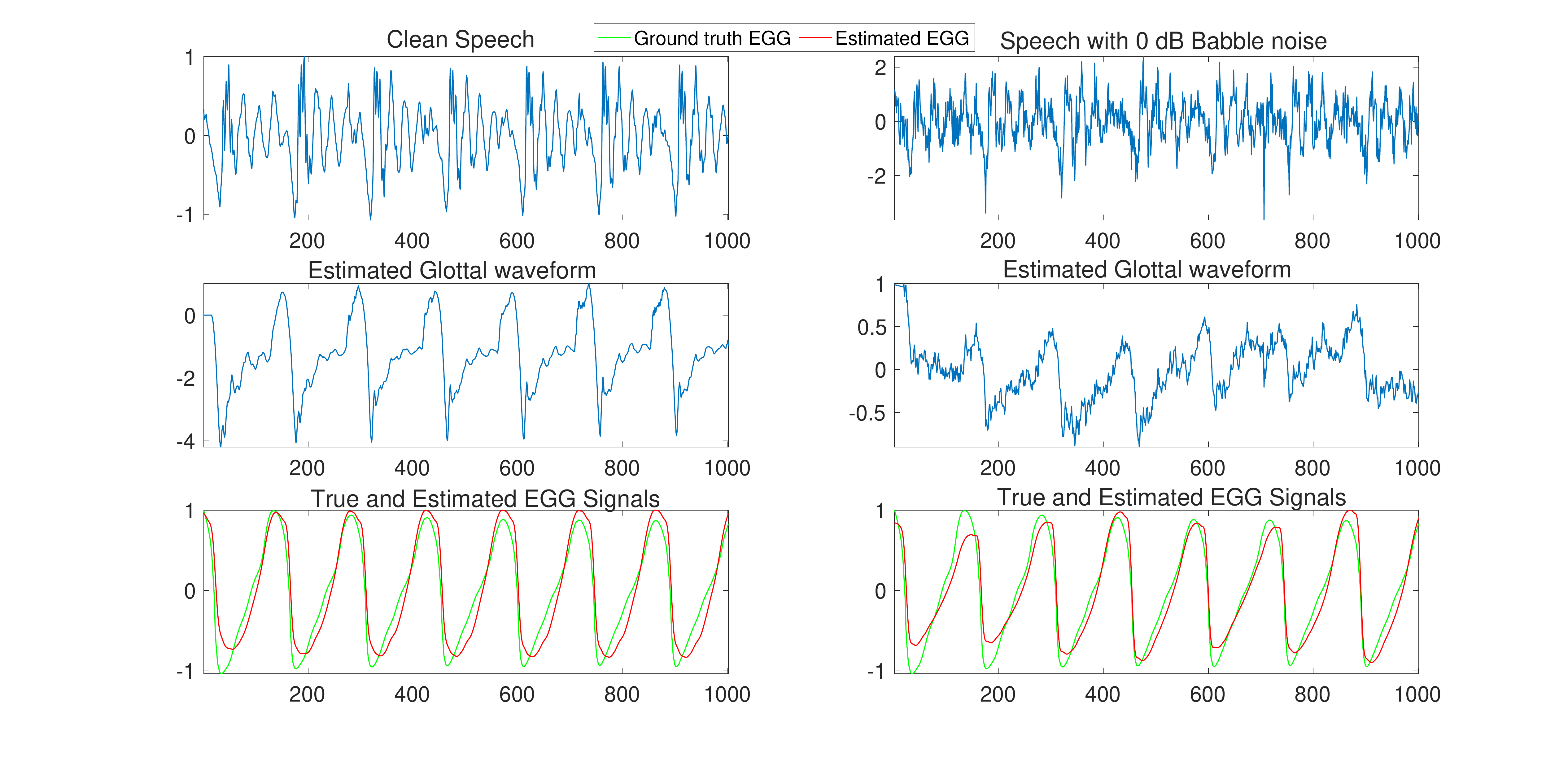}
	\centering
	\caption{Depiction of a model-based glottal source estimation technique - the left trace is a segment of voiced speech with corresponding glottal source estimated using linear prediction inverse filtering. Right trace is the same segment corrupted with additive Babble noise at 0 dB SNR. It can be seen that the estimated glottal waveform is heavily corrupted when the speech is noisy albeit the EGG signal is barely altered since it is independent of acoustic signal.}
	\label{fig:source_est}
\end{figure}

\subsection{Need for estimation of the EGG}
Even though EGG offers several advantages over other glottal estimation techniques, additional hardware is required for its extraction which might not be available, especially in non-clinical settings. Further, it is difficult to record EGG for people with thick neck tissues and is sensitive to the sporadic low-frequency body movements \cite{childers1985critical}. Thus it would useful if the EGG waveform could be estimated devoid of physical hardware. Many previous works exist to extract information about the parameters of the EGG waveform directly from the speech signal. The typical approach is to estimate the glottal flow waveform and estimate the glottal flow parameters from it \cite{alku2011glottal}.
Nevertheless, we hypothesize that the estimation of the the EGG waveform (and thus the glottal flow parameters) directly from the acoustic signals is desirable due to the following reasons:

\begin{enumerate}

	\item Several applications \cite{titze1990interpretation, peterson1994comparison,chen2002electroglottographic} of EGG demand the morphological analysis of the entire waveform rather than the summary parameters. Thus it is desirable to extract the entire EGG signal rather than parameters quantifying it.
	\item Even in the case when the application demands only the estimation of the glottal parameters, we hypothesize (and provide empirical evidence) that it is optimal to estimate it from the EGG signal rather than a glottal flow estimate. This is because of the inherent vulnerability associated with the parametric models of glottal flow estimation \cite{alku2003parameterisation} that suffer heavily if there are deviations from the model assumption, change in recording conditions (see Fig. 2 for an example) and varying voice characteristics (breathy voice, falsetto) \cite{drugman2012comparative} which is not present in the EGG waveform.
	\item Glottal flow estimators  are task-specific and demand a-priori information such as average pitch period \cite{murty2008epoch},  glottal closure instants \cite{drugman2008voice}, the region of voicing \cite{prathosh2013epoch} whereas a direct EGG estimator is devoid of all these rather can provide these information as  by-product.
	\item The availability of the true EGG signals simultaneously acquired with the corresponding speech signal makes it possible to devise an estimator for the EGG without the need for making a model assumption.
	\item There are dedicated algorithms proposed in the literature to extract glottal parameters (such as GCI, Open Quotient) from the estimated glottal waveform \cite{drugman2012detection,alku1992glottal}. However, estimating these from the EGG waveform is much simpler. For example, GCI could be detected trivially with a threshold on the differentiated version of the EGG signal.

\end{enumerate}

\subsection{Contributions}

Owing to the aforementioned need and the tremendous success of the data-driven neural models in distributional learning, we attempt to address the problem of directly estimating the Electroglottography signal from the speech signal. We approach the problem from a  distribution transformation perspective and employ the principles of the supervised variational inference to conditionally generate the EGG signal given a speech segment. To the best of our knowledge, this is the first attempt to  generate the whole EGG signal from speech signal. We leverage the abundant availability of the simultaneously recorded speech and EGG data for this task. If successful, this could replace the bulky and expensive EGG device and aid in many applications such as screening of speech disorders, voice quality assessment, pitch estimation etc. The following list briefly summarizes the contributions of this work.

\begin{enumerate}
	\item Formulation of the problem of speech to EGG conversion from a data-driven distribution transformation perspective.

	\item Introduction of a general method of approximate inference for conditional distribution transformation through optimization of the evidence lower bound constructed by minimizing KL divergence between the true and the approximate posteriors.

	\item Use of an informative prior derived from a neural autoencoder, that is known to reconstruct the EGG signal.

	\item Employing adversarial learning principles for imposition of the learned informative prior on the latent space of the distribution transformation network.


	\item Demonstration of the efficacy of the method for the speech to EGG conversion task through rigorous generalization experiments with several temporal and spectral metrics, on multiple datasets comprising different speakers, recording conditions, noise characteristics, voice qualities and speech pathologies.

\end{enumerate}

The rest of the paper is organized as follows: Sections \ref{sec:probform} and \ref{sec:aai} formulate the problem and develop the theory for the adversarial approximate inference. Section \ref{sec:elbosec} discusses the nuances of realizing the method using neural networks. Sections \ref{sec:impl}, \ref{sec:expdata} and \ref{sec:expmetrics} describe the experimental protocol, data and the metrics used for the assessment. This is followed by the discussion of results in Section \ref{sec:results} and concluding remarks in Section \ref{sec:conclusion}.


\section{Proposed Methodology}

The task of converting speech to EGG is posed as a distribution transformation problem where the goal is to learn a non-linear transformation to map samples from the speech distribution to the EGG distribution. The existence of such a transformation is motivated by the fact that the underlying physical phenomena that gives rise to both the speech and the EGG signal is the same.

Concretely, learning the transformation of speech to EGG is cast as the problem of learning the parameters $\bp$ that maximize the conditional probability distribution $p_{\bp}(Y|X)$ where $(Y)$ and $(X)$ represent the EGG and speech signals, respectively. While maximization of this unknown probability distribution is an intractable problem in general, recent advances in the field of deep learning have allowed construction of successful generative models for estimating and sampling from a probability distribution, improving upon many of the caveats presented by classical sampling techniques such as Markov Chain Monte Carlo (MCMC). A brief outline of two such methods has been given below.

\subsection{Background on Neural Generative models}

\subsubsection{Generative Adversarial Networks}

The Generative Adversarial Networks (GAN) framework by \cite{goodfellow2014generative} forms one of the most popular approaches to deep generative models which cast distribution learning as a minimax game. Their primary advantage over classical sampling techniques such as MCMC lies in their ability for single step generation of samples from a desired high-dimensional distribution instead of the computationally intensive repeated sampling in Markov Chains \cite{goodfellow2016nips}. Optimizing a GAN involves an adversarial game between two neural networks - a generator $G(\bz)$, and a discriminator $D(\bx)$, where the objective is to match the generator distribution $P_G(\bx)$ to the true data distribution $P_\mathcal{D}(\bx)$. The generator, $G$ learns an implicit density by transforming samples from a (known) prior distribution $\bz \sim P(\bz)$ to samples $G(\bz)$ from the generator distribution, while the discriminator $D(\bx)$ predicts the probability that $\bx$ belongs to the true data distribution. The discriminator acts like a classifier that aims to distinguish between samples from the true data distribution $P_\mathcal{D}(\bx)$ and the generator's distribution $P_G(\bx)$. The game consists of the generator trying to fool the discriminator into believing that the generator samples come from the true data distribution, while the discriminator tries to correctly distinguish between the two. Formally, the solution to the game is a Nash equilibrium of the following value function.

\begin{equation} \label{eq:gan}
	\begin{aligned} \min_G \max_D V(G, D) = & \EX_{\bx \sim P_\mathcal{D}(\bx)} \left[ \log D(\bx) \right] + \\ &\EX_{\bz \sim P(\bz)}\left[ \log (1 - D(G(\bz))) \right]
	\end{aligned}
\end{equation}

\subsubsection{Variational Autoencoders}

Variational Autoencoders \cite{kingma2013auto,chen2016variational} alongside GANs form the other most popular approach to deep generative modelling. Inspired by variational bayesian inference, VAEs form a directed, graphical model where the distributions of the random variables are parameterized by neural networks, and the latent variables are assumed to come from a tractable, explicitly computable density function (such as the standard normal distribution). The graphical model consists of a conditional distribution $p_{\bt}(\bx|\bz)$ over the observed variables, an approximate posterior over the latent variables $q_{\bp}(\bz | \bx)$ and a specified prior (with known density) $p(\bz)$. Then, it can be shown that \cite{kingma2013auto}:

\begin{equation} \label{eq:vae}
	\begin{aligned}
		\log p_{\bt}(\bx) & \ge -D_{KL}[q_{\bp}(\bz | \bx) || p_{\bt}(\bz)] + \EX_{q_{\bp}(\bz | \bx)} [\log p_{\bt}(\bx | \bz)]
	\end{aligned}
\end{equation}

where $\bt,\bp$ are parameters of a neural network. In VAEs, the equation \eqref{eq:vae} is used to optimize a lower bound to the likelihood, since the true likelihood $p_{\bt}(\bx)$ which requires marginalization over the latent variables $\int_{\bz} p_{\bt}(\bx,\bz) d \bz$ is usually intractable.

\subsection{Problem Formulation} \label{sec:probform}

The roots of the task of mapping the distribution of speech to EGG lie in learning the conditional distribution of EGG given speech. While GANs have had remarkable success in
sampling from an arbitrary data distribution, learning in a conditional setting is known to be unstable and notoriously hard to train \cite{goodfellow2014generative}. Secondly, GANs in their original formulation cannot incorporate supervised labels (pairs of speech and EGG segments in this case) into the training paradigm. However, several variations of GANs have been proposed both to stabilize the GAN training and impose a conditioning variable into  the generative model \cite{mirza2014conditional,gulrajani2017improved}. On the other hand, variational auto encoders (VAEs) while robust to training variations, make the assumption of putting a standard normal prior on the latent space. This assumption is not necessarily satisfied in practice, especially when the latent distribution is known to deviate from the standard normal distribution (e.g., a multi-modal distribution) \cite{burgess2018understanding}. Thus, it is desirable to impose such properties on the distribution of the latent space that aid the process of conditional generation. In this work, we aim to propose a stable conditional generative model that imposes an informative latent prior through adversarial learning, via the principles of the variational inference.
Given the immense variability of human utterances in terms of phonemes, co-articulation, speakers, voice types, gender etc., the distribution of speech samples is expected to have a very high entropy. In contrast to this, the distribution of EGG is expected to possess a lower entropy since it embeds very less information as compared to the speech signal. Further EGG is predominantly a low-pass signal \cite{childers1985critical} (Fig. \ref{fig:source_est}) without any formant information. This naturally suggests that there exists a lower dimensional representation $(Z)$ of the speech signal $(X)$ that is more amenable for extracting the EGG information while discarding the variations that arise due to spectral coloring by the vocal tract. We propose to exploit this structure in our formulation by constructing an information bottleneck, and enforcing learning of a lower dimensional representation, that is known to reconstruct the EGG signal.
The challenge with approaches trying to perform distribution transformation is the lack of explicit density functions for the source and the target distributions. However, one would have access to  noisy samples that are drawn from both of these distributions, which are the speech and the corresponding EGG signals in this case. Hence, we use an approach inspired by variational inference to provide a lower bound (an approximation) on the likelihood of the target distribution, which has a tractable form that is suitable for optimization.

\subsection{Adversarial Approximate Inference} \label{sec:aai}

\noindent In our supervised learning setting, the dataset takes the form
$\{ ( \bx_1 , \by_1 ), \cdots ,( \bx_N , \by_N ) \}$ where $( \bx_i , \by_i )$ is the \textit{i}-th observation consisting of a speech segment $\bx_i$ and its corresponding EGG signal $\by_i$. We assume existence of a lower dimensional representation or a continuous latent variable, $\bz$, of a speech segment $\bx$ that allows reconstruction of the corresponding EGG segment, $\by$.

Let $p_{\bt^*} ( \bz | \bx )$ denote the true posterior distribution over the latent variables conditioned on the speech samples. Since the true distribution is unknown, we propose to learn a parameterized approximation $q_{\bp} ( \bz | \bx )$ to the intractable true posterior $p_{\bt^*} ( \bz | \bx )$. Let $p_{\bt} (\bz | \by)$ denote the posterior on  the latent variable $\bz$ conditioned on the EGG samples $\by$. We assume that EGG can be perfectly reconstructed from $\bz$ (the latent space constructed from the EGG signals) since it is known that the distribution of EGG signal has a lower entropy than speech and thus can be learnt with low learning complexity. Then it is intuitive to map the input speech samples to such a latent space that would reconstruct the EGG well.  This can be achieved by minimizing the KL divergence between the EGG conditional distribution, $p_{\bt} (\bz | \by)$ and the approximation $q_{\bp} (\bz | \bx)$ to the  true posterior $p_{\bt^*} ( \bz | \bx )$, that transforms speech to the EGG signal. Mathematically,

\begin{equation}
	\begin{aligned}
		 & D_{KL}[q_{\bp} ( \bz | \bxi ) \, || \, p_{\bt} (\bz | \byi)]                              \\
		 & = \EX_{q_{\bp}} [ \log q_{\bp} ( \bz | \bxi ) ] - \EX_{q_{\bp}}[ \log p_{\bt}(\bz|\byi) ]
	\end{aligned}
\end{equation}

\begin{equation}
	\begin{aligned}
		 & \Rightarrow D_{KL}[q_{\bp} ( \bz | \bxi ) \, || \, p_{\bt} (\bz | \byi)] \\
		 & =  \EX_{q_{\bp}} [ \log q_{\bp} ( \bz | \bxi ) ]
		- \EX_{q_{\bp}} [ \log p_{\bt} ( \byi | \bz ) ]                             \\
		 & - \EX_{q_{\bp}} [ \log p_{\bt} (\bz) ]
		+ \EX_{q_{\bp}} [ \log p_{\bt} (\byi) ]
	\end{aligned}
\end{equation}

Since the distribution over $p_{\bt}(\byi)$ does not depend on $q_{\bp} ( \bz | \bxi )$, we can write the marginal log-likelihood of the EGG distribution as

\begin{equation}
	\begin{aligned}
		\log p_{\bt} (\byi)
		=  D_{KL}[q_{\bp} ( \bz | \bxi ) \, || \, p_{\bt} (\bz | \byi)] +
		\mathcal{L}( \bp, \bt ; \bxi)
	\end{aligned}
\end{equation}

where the lower bound on the log-likelihood $\mathcal{L}( \bp, \bt ; \bxi)$ takes the form

\begin{equation} \label{eq:elbo}
	\begin{aligned}
		\mathcal{L}( \bp, \bt ; \bxi) = & - D_{KL}[q_{\bp} ( \bz | \bxi ) \, || \, p_{\bt} (\bz )] \\ &+ \EX_{q_{\bp}} [ \log p_{\bt} (\byi | \bz) ]
	\end{aligned}
\end{equation}

As $ D_{KL}[q_{\bp} ( \bz | \bxi ) \, || \, p_{\bt} (\bz | \byi)] \ge 0$, we have the following inequality

\begin{equation}
	\begin{aligned}
		\log p_{\bt} (\byi)                             & \ge \mathcal{L}( \bp, \bt ; \bxi)                      \\
		\log p_{\bt} (\byi | \bxi) + \log p_{\bt}(\bxi) & \ge \mathcal{L}( \bp, \bt ; \bxi)                      \\
		\log p_{\bt} (\byi | \bxi)                      & \ge \mathcal{L}( \bp, \bt ; \bxi) - \log p_{\bt}(\bxi)
	\end{aligned}
\end{equation}

As $- \log p_{\bt}(\bxi) \ge 0$, the evidence lower bound $\mathcal{L}( \bp, \bt ; \bxi)$ becomes a lower bound on the conditional distribution $p_{\bt} (\byi | \bxi)$.

Adopting a maximum likelihood based approach, we optimize this evidence lower bound $\mathcal{L}( \bp, \bt ; \bxi)$ (ELBO) with respect to the 'variational' parameters $\bp$ to learn the approximate distribution. If the true posterior distribution were known (i.e. $p_{{\bt}^*} (\bz | \byi)$), then the optimization problem would reduce to $\max_{\bp} \mathcal{L}( \bp, {\bt}^* ; \bxi)$. However, in practice, since the true posterior is unknown, it becomes a joint optimization problem over the variational and generative parameters $\{\bp,\bt\}$ respectively. Hence, the final optimization problem can be stated as

\begin{equation}
	\begin{aligned}
		\max_{\bp,\bt} \frac{1}{N}\sum_{i=1}^{N}\mathcal{L}( \bp, \bt ; \bxi)
	\end{aligned}
\end{equation}

\begin{figure*}[h!]
	\centering
	\includegraphics[trim={0 0 1cm 0},clip,width=\textwidth,height=6.5cm]{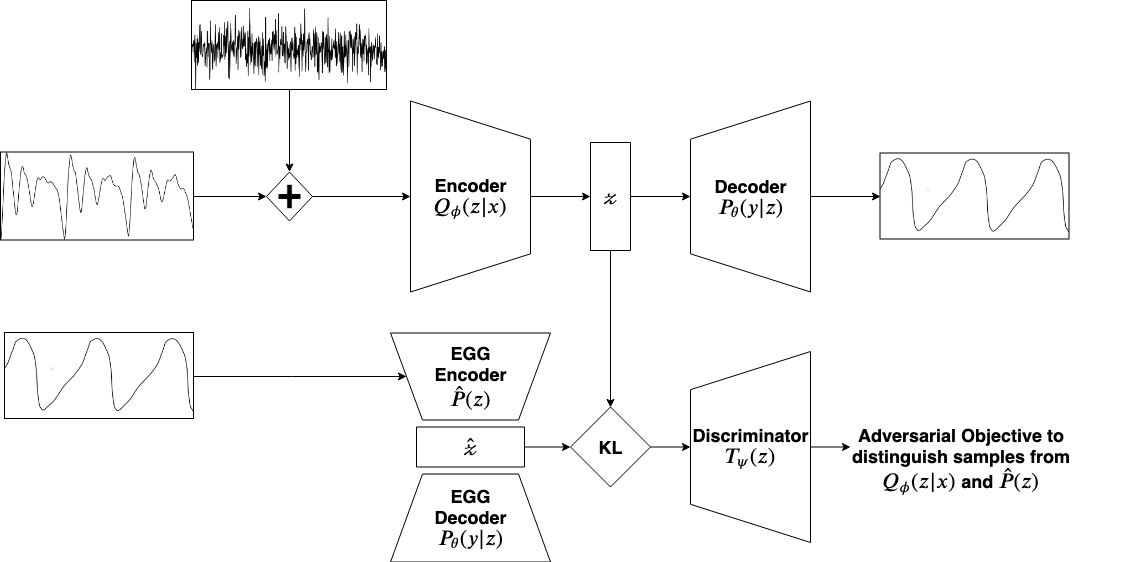}

	\caption{Architectural depiction of  the proposed method realized using neural networks. The encoder $Q_{\bp}$ tries to learn the encoding ($\bz$) from the speech to the EGG space, specified by the EGG encoder $P_{\bt}(\bz)$. This is facilitated through an adversarial training with the discriminator $T_\psi$. Finally, the decoder $P_{\bt}(\by | \bz)$ remaps $\bz$ to the EGG space to generate the corresponding EGG signal.}

	\label{fig:nnc}
\end{figure*}

\begin{algorithm}
	\caption{Adversarial Approximate Inference (AAI)}
	\label{aai}
	\begin{algorithmic} 
		\INPUT{Dataset $\mathcal{D}$, Generative Model $p_{\bt}(y|z)$, Transformation Model $q_{\bp}(z | x)$, Discriminator $T_{\bs}(z)$, Prior $p_{\bt}(z)$, Noise Distribution $p_{\epsilon}(\epsilon)$, Iterations $K$, Batchsize $B$}

		\Repeat
		\For{$k = 1 \; \mathbf{to} \; K$}

		\State Sample $\{x^{(1)} \cdots x^{(B)} \}$ from dataset $\mathcal{D}$
		\State Sample $\{\epsilon^{(1)} \cdots \epsilon^{(B)} \}$ from $p_\epsilon(\epsilon)$

		\State $x^{(i)} \gets x^{(i)} + \epsilon^{(i)}$
		\State $z_Q^{(i)} \gets$ Sample from $q_{\bp}(z | x^{(i)})$

		\State Compute gradient with respect to $\bt$
		\State \vspace{-0.3cm} $$\bm{g}_{\bt} \gets \frac{1}{B} \sum_{j=1}^{B} \nabla_{\bt} \left[ \log p_{\bt}(y^{(j)} | z_Q^{(j)})  \right]$$
		\State Compute gradient with respect to $\bp$
		\State \vspace{-0.3cm} \begin{align*}
			\bm{g}_{\bp} \gets \frac{1}{B} \sum_{j=1}^{B} \nabla_{\bp} \Big[
			 & - \log (1 - T_{\bs}(z_Q^{(j)}))    \\
			 & +\log p_{\bt}(y^{(j)} | z_Q^{(j)})
			\Big]
		\end{align*}

		\State Update $\bt, \bp$ using gradients $\bm{g}_{\bt}, \bm{g}_{\bp}$
		\EndFor

		\State Sample $\{x^{(1)} \cdots x^{(B)} \}$ from dataset $\mathcal{D}$
		\State Sample $\{\epsilon^{(1)} \cdots \epsilon^{(B)} \}$ from $p_{\epsilon}(\epsilon)$

		\State Sample $\{z_p^{(1)} \cdots z_p^{(B)} \}$ from $p_{\bt}(z)$
		\State $x^{(i)} \gets x^{(i)} + \epsilon^{(i)}$
		\State $z_Q^{(i)} \gets$ Sample from $q_{\bp}(z | x^{(i)})$

		\State Compute gradient with respect to $\bs$
		\State \vspace{-0.3cm}
		\begin{align*}
			\bm{g}_{\bs} \gets \frac{1}{B} \sum_{j=1}^{B} \nabla_{\bs} \Big[
			 & + \log (1 - T_{\bs}(z_Q^{(j)})) \\
			 & +\log  T_{\bs}(z_p^{(j)})
			\Big]
		\end{align*}

		\State Update $\bs$ using gradients $\bm{g}_{\bs}$

		\Until{convergence of $\bt, \bp, \bs$}
	\end{algorithmic}
\end{algorithm}

\subsection{Optimizing the ELBO} \label{sec:elbosec}

While any optimization technique will achieve a lower bound on the log-likelihood of the EGG by optimizing $\mathcal{L}$, the quality of the model is critically dependent on the tightness of the variational bound, as well as the assumption that the true data distribution is well approximated by $p_{\bt^{*}}(\by | \bz)$.

The parametrization chosen for the inference or recognition model $q_{\bp}( \bz | \bxi)$ naturally decides the tightness of the lower bound derived above. If there exists $\bp^* \in \bm{\Phi} \; \text{s.t.} \; q_{\bp}(Z|X) = p_{\bt}(Z|X)$ where $\bm{\Phi}$ is the space of inference parameters, then the ELBO will be a tight bound to $\log p_{\bt} (\byi)$. Unfortunately, this rarely happens in practice, hence, the space $\bm{\Phi}$ is designed to be as expressive as possible, to allow learning a close approximation. In our model, we parameterize both $q_{\bp}( \bz | \bxi), \; p_{\bt} ( \byi | \bz )$ as neural networks which are known to be universal function approximators. As pointed out in \cite{kingma2014semi}, this is much more efficient than the classical approach of Variational EM maximization where the joint optimization would involve separate parameters for each data point instead of shared global parameters in a neural network.


In the original formulation of Variational Autoencoders \cite{kingma2013auto,rezende2014stochastic}, the prior $p_{\bt} (\bz )$ is chosen to be the distribution $\mathcal{N}(\bm{0}, \mathbb{I})$ for ease of sampling, tractability and to obtain a closed form expression for the objective function. This limits both the representation capacity of the learnt variational parameters $\bp$ as they are restricted to be close to an arbitrary distribution, and may cause severe underfitting in the worst case. A weak non-representative prior further exacerbates the underfitting problem in a variational setting, as too weak a prior, will lead to very weak variational bound even in the limit of infinite data and perfect optimization \cite{goodfellow2016nips}.
Thus, choosing a \textit{good} prior is crucial in a variational setting. We address these caveats by building an end to end differentiable model where the distribution $p_{\bt}(\bz)$ is learnt by a separate autoencoder, known to perfectly reproduce the EGG signal and ensure that the prior over the latent space is informative enough to achieve tight bound on the log-likelihood. An autoencoder that reconstructs the EGG signal imposes a latent space $\bz$ and learns a marginal distribution over it given by the following equation

\begin{equation}
	\begin{aligned}
		p(\bz)=\int_y p(\bz | \by) p(\by) d \by
	\end{aligned}
\end{equation}

If one assumes a perfect reconstruction of the EGG by the EGG autoencoder (as validated by our experiments), this marginal distribution can be considered a \textit{good} prior, since it is known to allow EGG reconstruction from the latent space. In all future exposition, we refer to this marginal distribution as $p_{\bt}(\bz)$, under the assumption that a low empirical reconstruction loss, implies it is close to the optimal prior. Once this prior is learnt through the EGG autoencoder, we enforce learning of the same distribution at the latent layer of the network that converts speech to EGG. This is done by adversarially minimzing the KL divergence $D_{KL}[q_{\bp} ( \bz | \bxi )\, || \, p_{\bt} (\bz )]$ term in equation \eqref{eq:elbo}.
As shown in \cite{makhzani2015adversarial}, adversarial training can be used to train $q_{\bp}(\bz|\bxi)$ to be a universal approximation of the posterior, by augmenting the input $\bxi$ with random noise $\epsilon$.
This allows construction of arbitrary posteriors $q_{\bp}(\bz|\bxi)$, by evaluating the inference model $f_{\bp}(\bxi, \bm{\epsilon})$ for different values of $\epsilon$. Thus, even with a fixed deterministic mapping from input to the latent space, the posterior will not collapse to a degenerate Dirac delta distribution (i.e. a discontinuous distribution), as the input noise adds a source of stochasticity other than the data generating distribution itself. Hence, the actual posterior is given by the expression

\begin{equation}
	\begin{aligned}
		q_{\bp}(\bz|\bxi) = \int_{\epsilon} q_{\bp}(\bz|\bxi, \bm{\epsilon}) p_\epsilon(\epsilon) d \epsilon
	\end{aligned}
\end{equation}

where $q_{\bp}(\bz|\bxi, \bm{\epsilon})$ is the degenerate distribution $\delta(\bz - f_{\bp}(\bxi, \bm{\epsilon}))$. This fact is corroborated by experiments, where it is seen that inducing noise in the training data generalizes better. In addition, our method removes the normality constraints both on the posterior as well as the prior distribution by using adversarial training to minimize the KL divergence. This would involve a minimax game between the $q_{\bp}(\bz | \bx)$ network and a discriminator network $T_{\bs}(\bz)$, that is poised to detect whether the sample given by the $q_{\bp}(\bz | \bx)$ network comes from the prior $p_{\bt}(\bz)$ (learned through the EGG autoencoder) or not. Mathematically, the following objective function is optimized to minimize the KL term in ELBO.

\begin{equation} \label{eq:gan}
	\begin{aligned} \min_{q_{\bp}} \max_{T_{\bs}} V(q_{\bp}, T_{\bs}) = & \EX_{\bz \sim p_{\bt}(\bz)} \left[ \log T_\psi(\bz) \right] + \\ &\EX_{\hat{\bz} \sim q_{\bp}(\bz|\bx)}\left[ \log (1 - T_\psi(\hat{\bz})) \right]
	\end{aligned}
\end{equation}

The second term of the ELBO $\mathcal{L}( \bp, \bt ; \bxi)$ (Eq. \eqref{eq:elbo}) is interpreted as the expected EGG reconstruction error at the output given the latent vector, and can be minimized by a number of loss functions. We choose to minimize this error by measuring the cosine distance between the estimated and the ground truth EGG to impose stronger restrictions on the shape of the learnt EGG which is a defining characteristic of the signal. The long-term amplitude of the EGG is known to be an artifact of the measurement apparatus and use of the cosine loss is expected to aid invariance to the superfluous variations in amplitude under different recording and environmental settings. Mathematically, if $\hat{y}$ and $y$ denote the estimated and the true EGG respectively, the cosine distance loss $\mathcal{L}(\hat{\by}, \by)$ is given by

\begin{equation}
	\mathcal{L}(\hat{\by}, \by) = cos^{-1} { \left( \frac{\langle \hat{\by}, \by \rangle}{||\hat{\by}|| \hspace{1mm} ||\by||} \right)}
\end{equation}

\begin{figure}[ht!]
	\centering
	\includegraphics[width=0.45\textwidth,height=8cm]{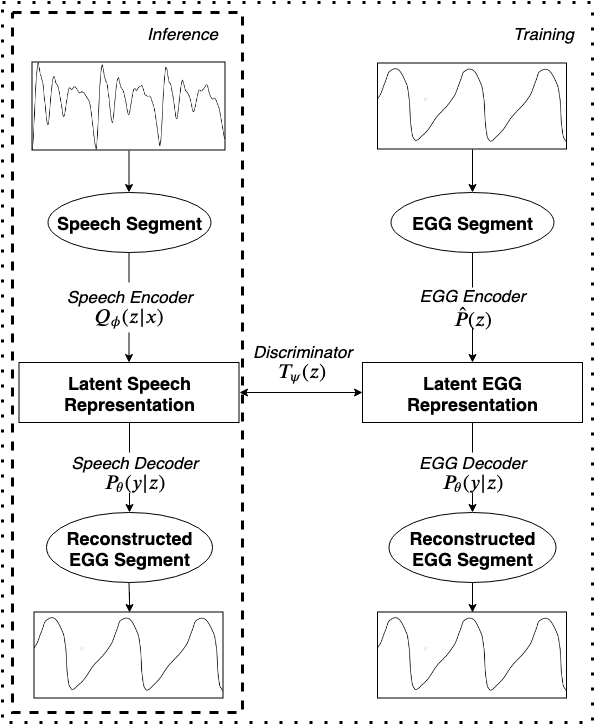}
	\caption{A flowchart of the training and inference procedure on AAI. The training procedure consists of (i) training an EGG autoencoder (shown on the right) and (ii) training a speech to EGG encoder-decoder network where the encoder is adversarially trained to produce representations similar to the representations learnt by the EGG autoencoder. The inference procedure consists of passing a speech sample through the speech to EGG network shown on the left, which outputs the corresponding EGG signal.}

	\label{fig:flowchart}
\end{figure}

Since we are using (i) the principles of variational inference to optimize an approximation (lower bound) to the true likelihood, (ii) the principles of adversarial training to impose an informative prior on the latent space of the speech to EGG transformer, we name our method Adversarial Approximate Inference (AAI) whose summary is depicted in Fig. \ref{fig:nnc} and flow-chart in Fig. \ref{fig:flowchart}.

\subsection{Implementation Details} \label{sec:impl}

\begin{figure}[h!]
	\includegraphics[width=0.45\textwidth]{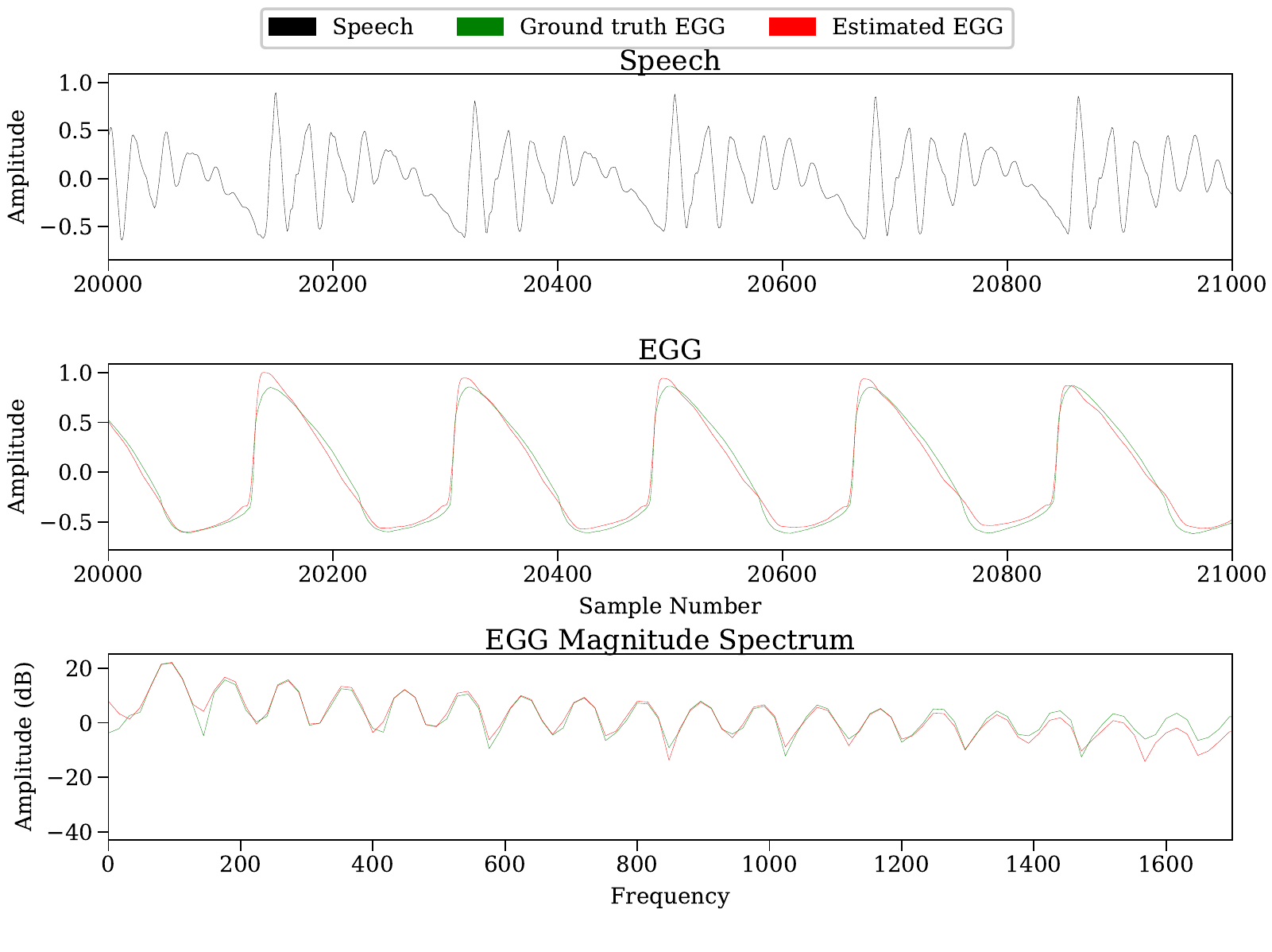}
	\centering
	\caption{Illustration of the AAI algorithm on a segment of a voiced speech with the corresponding Fourier spectra. It is seen that true and the estimated EGGs agree very well with each other both in terms of time and frequency domain characteristics.}
	\label{fig:hnr_spectra}
\end{figure}

\begin{figure}[h!]
	\includegraphics[width=0.45\textwidth, height = 0.3\textwidth]{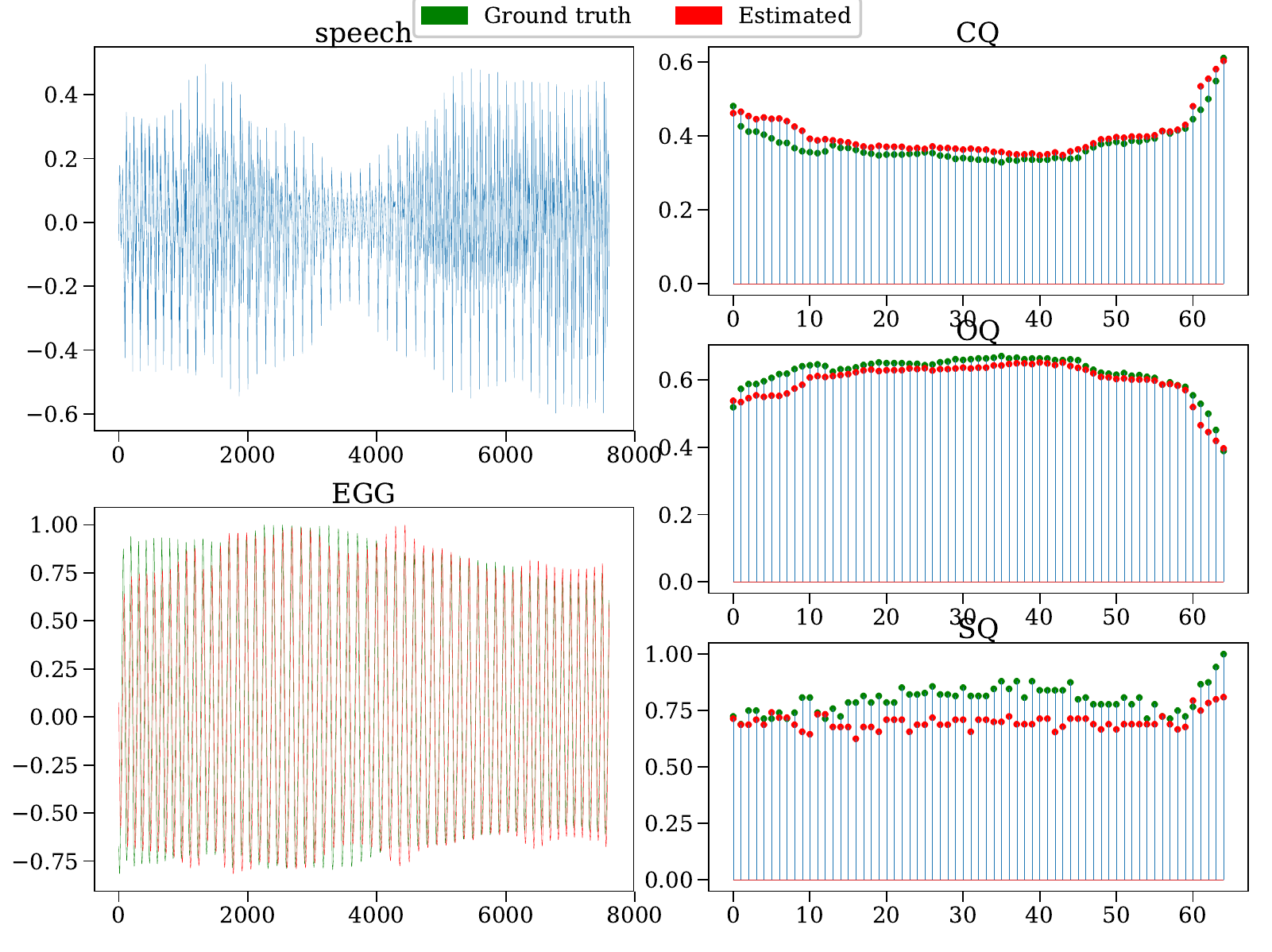}
	\centering
	\caption{Illustration of the AAI algorithm for speech to EGG conversion: A long segment of voiced speech with multiple phonemes is shown with the corresponding true and the estimated EGG signals. It is seen that the true and the estimated EGGs align closely with each other with the corresponding quotients.}
	\label{fig:cq_oq_sq}
\end{figure}

Given an input pair of speech and the corresponding EGG signals, we extract data points by framing them into 12 millisecond windows (the method would perform equally well with other window lengths as long as they have one or two pitch periods, on an average) with a stride length of a single sample. During inference, the predictions over overlapping windows are averaged and concatenated to obtain the final estimates. All the models including the EGG autoencoder, $q_{\bp}$ and $p_{\bt}$ networks are realized by fully connected neural networks of six layers with diminishing neurons interspersed with batch normalization layers. We employ the standard numerical optimization techniques such as stochastic gradient descent to learn all neural network parameters. The EGG autoencoder is first trained independently which is followed by training the $q_{\bp}$ and $p_{\bt}$ networks by employing the latent vectors generated by the trained EGG autoencoder. The complete algorithmic procedure of AAI is presented in Algorithm \ref{aai}. Figure \ref{fig:hnr_spectra} illustrates the performance of the AAI algorithm on a segment of speech signal. It is seen that the true and the estimated EGGs agree very well with each other both in terms of time and frequency domain characteristics. Figure \ref{fig:cq_oq_sq} depicts a long segment of voiced speech with multiple phonemes with the corresponding true and the estimated EGG signals. It is seen that the true and the estimated EGGs align closely with each other with the corresponding quotients.


\section{Experimental Setup} \label{sec:exp}

\subsection{Dataset description} \label{sec:expdata}

The effectiveness of the proposed methodology is demonstrated on multiple tasks and datasets with different speakers, voice qualities, multiple languages as well as speech pathologies. All datasets used for learning and evaluation purposes consist of simultaneous recordings of speech and the corresponding EGGs. The generated EGGs are evaluated on several metrics to ascertain their quality, compared to the ground truth EGGs. Parameters of all the networks are learnt on data provided in the book by D.G. Childers \cite{childers2000speech}, referred to as Childers' data. This dataset consists of simultaneous recordings of the speech and the EGG signals from 52 speakers (males and females) recorded in a single wall sound room. Childers' Data consist utterances of 16 fricatives, 12 vowels, digit counting from one to ten with increasing loudness, three sentences and uttering 'la' in a singing voice. For assessing the generalization, the learnt model has been tested on three datasets as described below.

\begin{itemize}
	\item \textbf{CMU ARCTIC} databases \cite{kominek2004cmu} consisting of 3 speakers SLT (\textit{US female}), JMK \textit{(Canada male)} and BDL \textit{(US male)} that has phonetically balanced sentences with simultaneously recorded EGGs.
	\item \textbf{Voice Quality (VQ)} database \cite{liu2017comparison} consists of recordings from 20 female and 20 male healthy speakers. The subjects phonated at conversational pitch and loudness on the vowel \textit{/a:/} in three ways, (i) habitual voice, (ii) breathy voice and (iii) pressed voice.
	\item \textbf{Saarbruecken Voice} database \cite{woldert2007saarbruecken}, referred to as \textit{Pathology} database, a collection of voice recordings from 2000 people both healthy and afflicted with several speech pathologies. The dataset contains recordings of vowels at various pitch levels as well as a recording of the sentence "Guten Morgen, wie geht es Ihnen?" (\textit{"Good morning, how are you?"}).
\end{itemize}

\begin{table}[]
	\centering{
		\caption{summary of the Datasets used in the study. All models are learnt on the Childers dataset and test on rest.}
		\label{tab:dataset-table}
		\begin{tabular}{@{}ccc@{}}
			\toprule
			Dataset   & No. Utterances & No. Glottal Cycles \\ \midrule
			Childers  & 200            & 1339302            \\
			CMU       & 3376           & 10338645           \\
			VQ        & 120            & 6926030            \\
			Pathology & 1873           & 376905             \\ \bottomrule
		\end{tabular}%
	}
\end{table}



\subsection{Evaluation Metrics} \label{sec:expmetrics}

Different assessment criteria are for evaluating the performance of the proposed method. As mentioned briefly in the introduction, EGG signal is characterized through multiple metrics that signify the shape of the EGG signal (and its derivative). Most popular metrics that are employed in the literature are the Glottal Closure Instants (GCI), Glottal Opening instants(GOI), Contact Quotient (CQ), Open Quotient (OQ), Speed/Skew Quotient (SQ) and Harmonic to Noise ratio (HNR). All the metrics are computed on the ground truth and the estimated EGGs using the same procedure. In the following section, we discuss each of these metrics in detail (Refer Fig. 1 for a pictorial depiction).


\begin{enumerate}
	\item GCI Detection: The instant of significant excitation in a single pitch period is defined to be an Epoch which coincides with the instant of maximal closure of the glottis (GCI) \cite{ananthapadmanabha1979epoch}. GCIs manifest as significant negative peaks in the differentiated EGG (dEGG) signal (Fig. 1).  GCIs are considered to be one of the most important events worthy of detection \cite{drugman2012detection}. Since, our method directly transforms the speech into the corresponding EGG signal, it is devoid of the requirement for the use of auxiliary signals (such as Voice source estimate) to detect GCIs. Further, since GCIs are present only during the voiced speech, most of the state-of-the-art GCI detectors rely on voiced-unvoiced classification as a necessary first step. However, AAI does not demand for a-priori voiced-unvoiced classification as it fully estimates the EGG. We extract GCIs (both true and the estimated EGGs) by picking the negative peaks in the dEGG signal using standard peak-picking algorithms. The de-facto metrics for the GCI detection task, namely, Identification Rate (IDR - \% of correct detections), Miss Rate (MR - \% of missed detections), False Alarm Rate (FAR - \% of false insertions) and Identification Accuracy (IDA - standard deviation of the errors between the predicted and the true GCIs) \cite{drugman2012detection} are used for evaluation and comparison purposes.
	\item GOI Detection Task: The complementary task of the GCI detection is the detection of instant of glottal opening. These manifest as positive peaks in dEGG signal (Fig. 1), are usually feeble in magnitude and very susceptible to noise. The same metrics used in the GCI detection problem are used to characterize the performance of GOI detection task as well.
	\item CQ: The contact quotient measures the relative contact time or the ratio of the contact time and time period of a single cycle \cite{liu2017comparison}.
	\item OQ: The open quotient measures the relative open time or the ratio of the period where the glottis is open to the time period of a single cycle.
	\item SQ: The speed quotient measures the ratio of the glottal opening time to the glottal closing time, and characterizes the degree of asymmetry in each cycle of the EGG signal. Both the SQ and the CQ have been observed to be sensitive to abnormalities in the glottal flow \cite{hillman1989objective}.
	\item HNR: The Harmonic to Noise ratio measures the periodicity of the EGG signal by decomposing the EGG signal into a periodic and an additive noise component. It provides a measure of similarity in the spectral domain, by using a frequency domain representation to compare the ratio of energy of the periodic component to the noise component. It is also known to quantify the hoarseness of the voice \cite{teixeira2013vocal}.
\end{enumerate}

The set of quotient metrics, i.e. CQ, OQ, SQ and HNR, both reference and estimated values are computed for every cycle in the ground truth and the estimated EGG and the summary statistics of the true and the estimated values are compared dataset-wise. For GCI detection task, we use the CMU Arctic datasets corrupted with two noise types, stationary white noise and non-stationary Babble noise at five different SNRs till 0 dB at steps of 5 dB. The same model that is trained for clean speech of the Childers' dataset is used for inference in the noise case as well. Further, we compare the proposed algorithm with five baseline state-of-the-art algorithms namely DPI \cite{prathosh2013epoch}, ZFR \cite{murty2008epoch}, MMF \cite{khanagha2014detection} and SEDREAMS \cite{drugman2009glottal}.

\begin{figure}[h!]
	\includegraphics[width=0.45\textwidth]{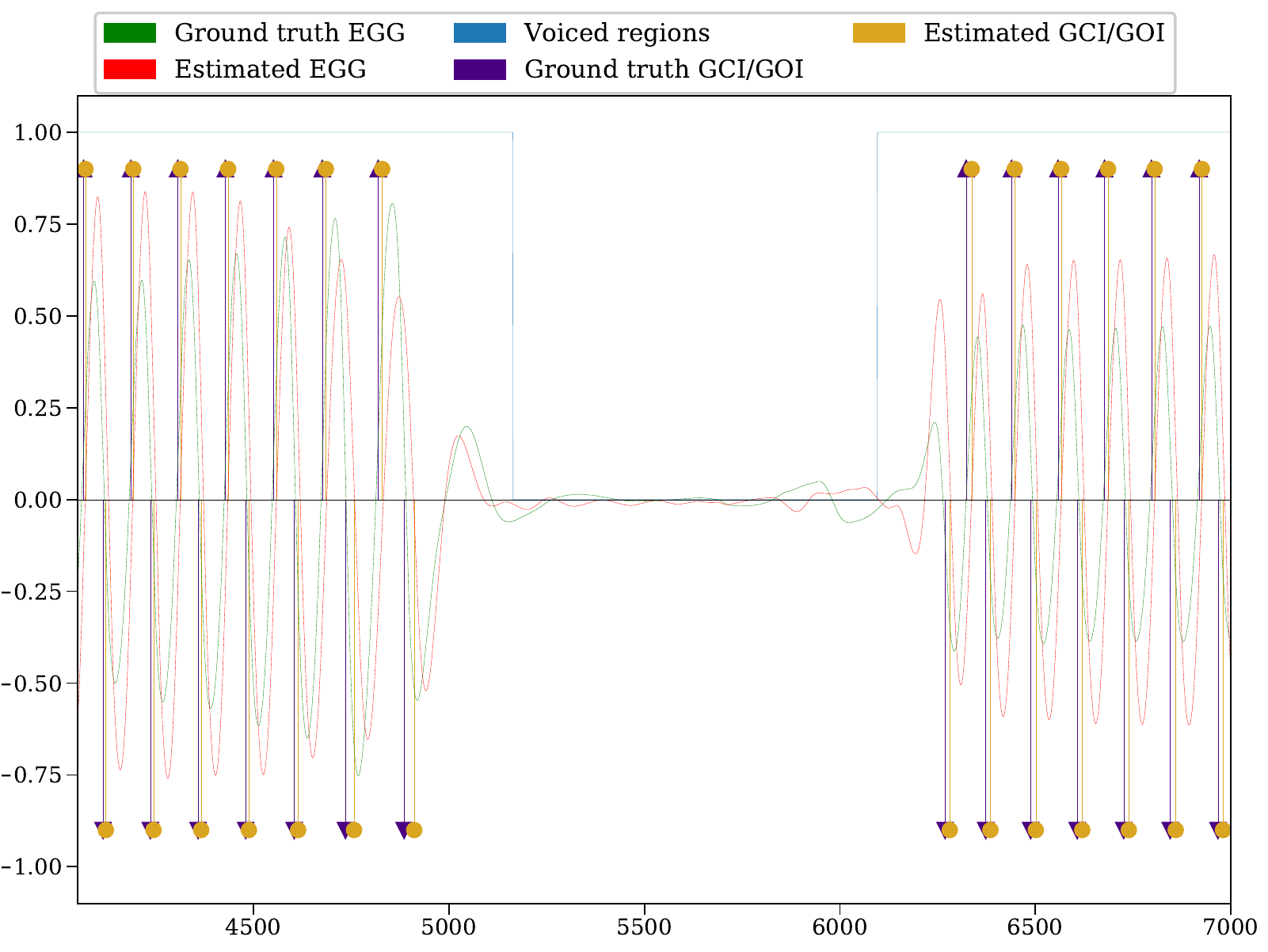}
	\centering
	\caption{Illustration of the AAI algorithm for GCI and GOI extraction on a segment of speech with a voiced-unvoiced boundary. It is seen that the GCIs and GOIs of the true and the estimated EGGs closely align with each other. Further, it can be observed that  the AAI algorithm identifies the boundary of the voicing as seen by the lower amplitude at the unvoiced regions.}
	\label{fig:vuv}
\end{figure}

\begin{figure*}[h!]
	\includegraphics[width=\textwidth, height=3.2 in]{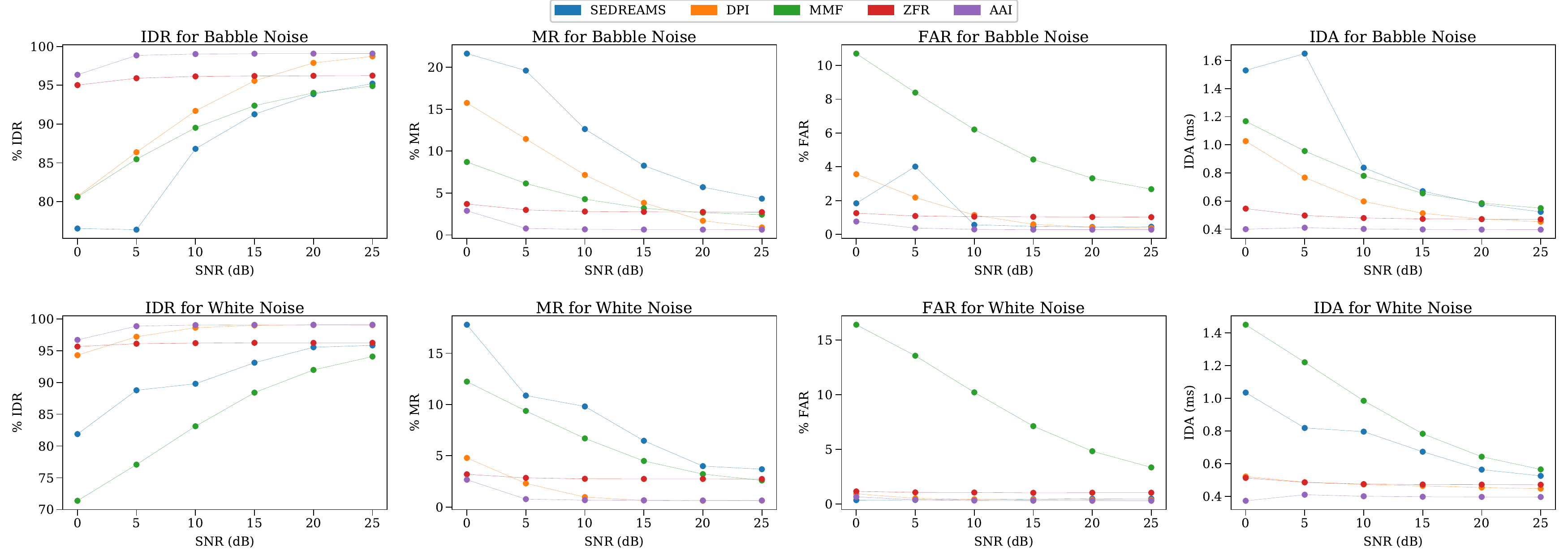}
	\centering
	\caption{Comparison of five state-of-the-art GCI detectors for white and babble noise at levels, on the CMU Arctic datasets.}
	\label{fig:noise5}
\end{figure*}

\begin{figure}[h!]
	\includegraphics[width=0.45\textwidth]{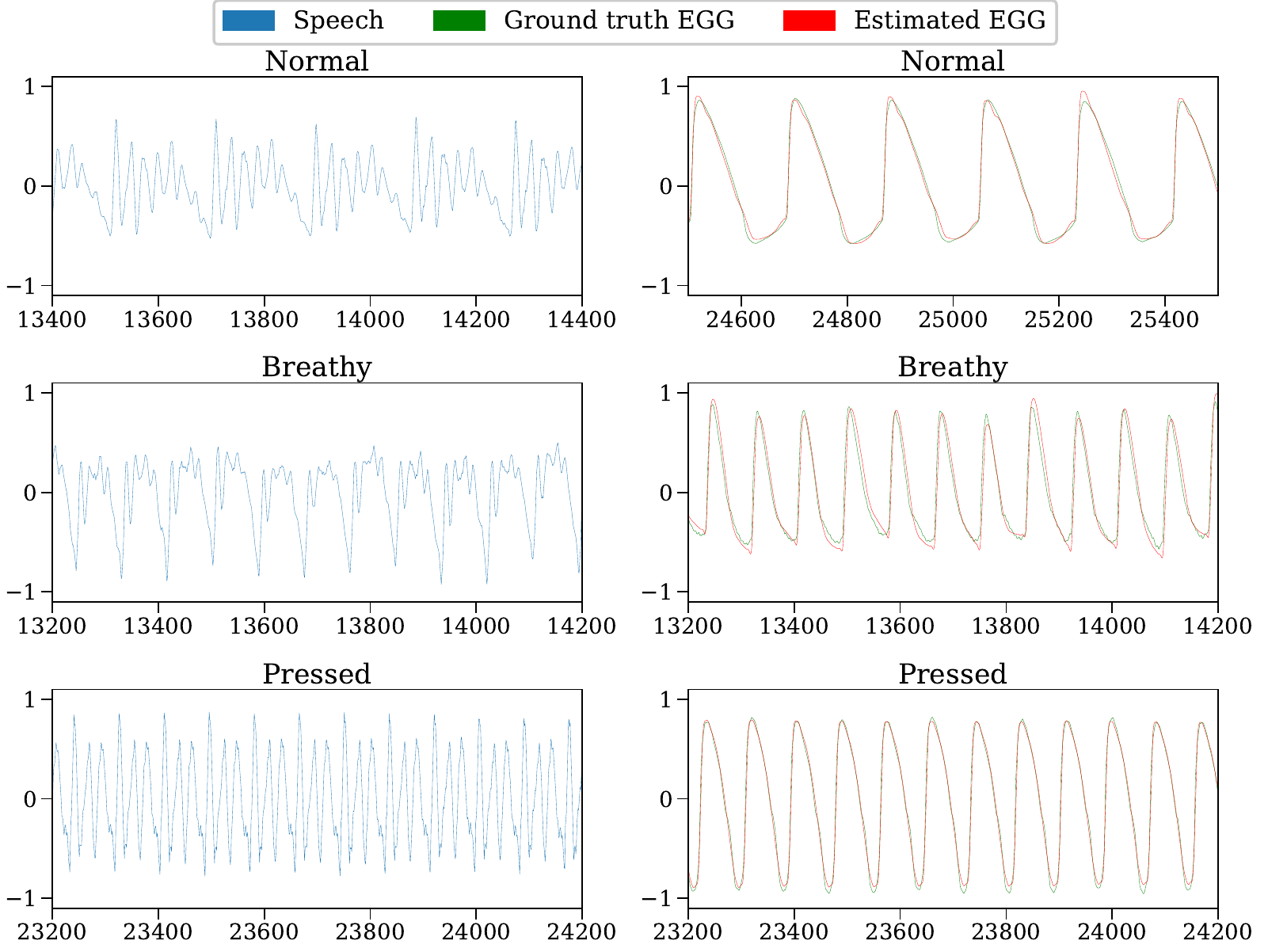}
	\centering
	\caption{Illustration of the AAI algorithm on speech samples of three different voice qualities.}
	\label{fig:voice_quality}
\end{figure}

\begin{figure}[h!]
	\includegraphics[width=0.45\textwidth]{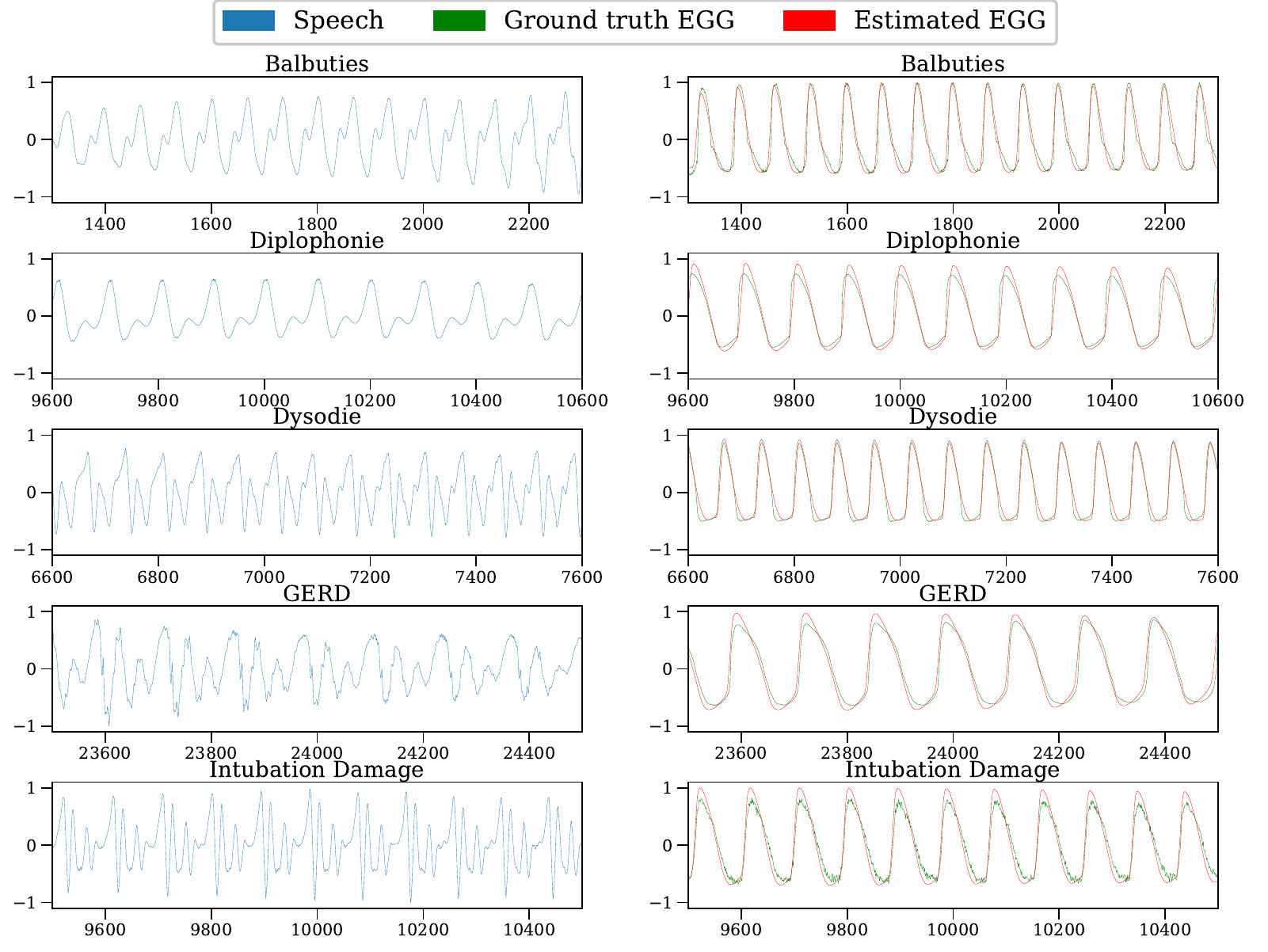}
	\centering
	\caption{Illustration of the AAI algorithm on speech samples of five different pathology types.}
	\label{fig:pathology}
\end{figure}

\begin{figure*}[h!]
	\includegraphics[width=\textwidth,height=4 in]{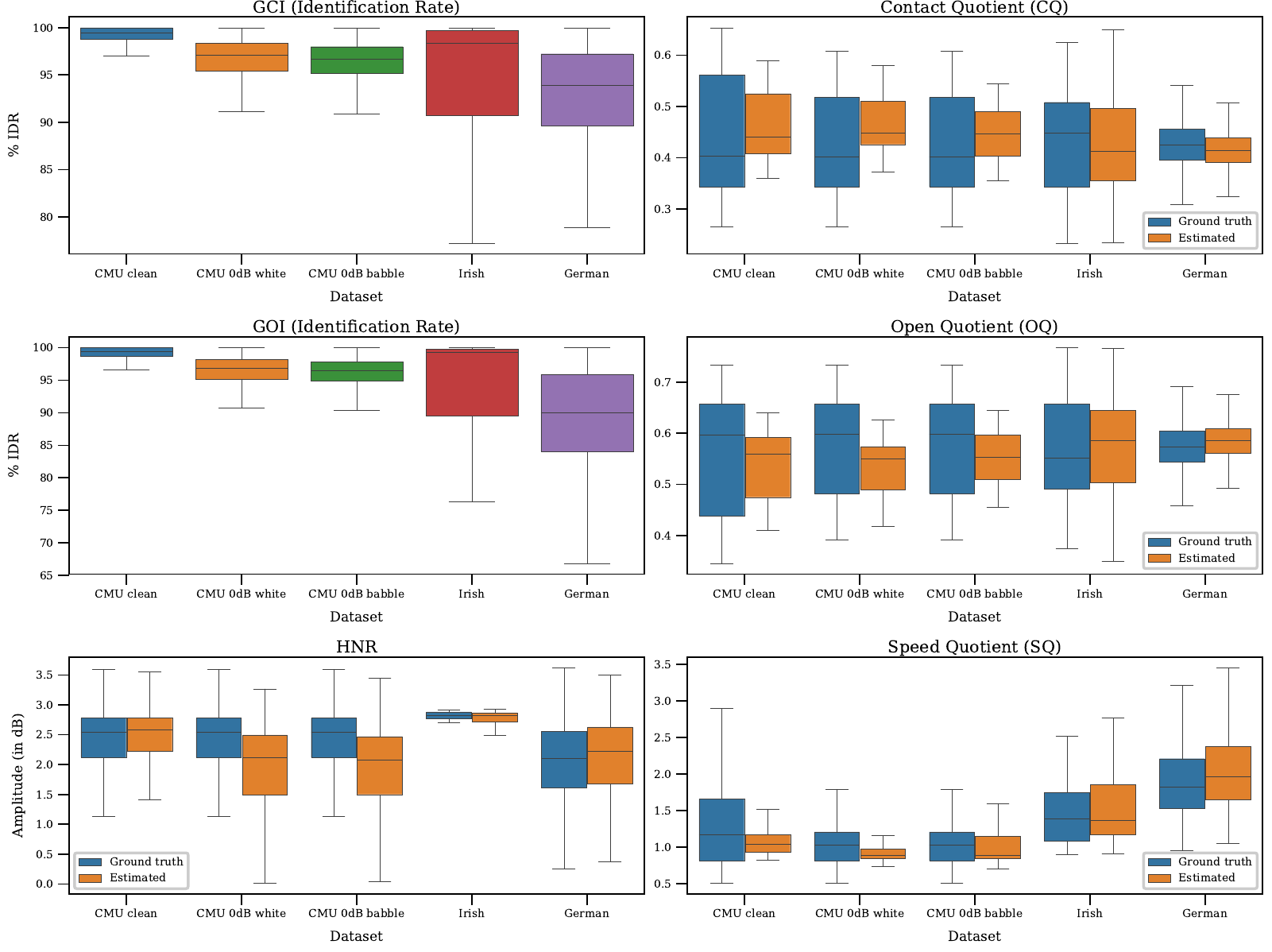}
	\centering
	\caption{Box-and-whisker plot for the entire set of metrics and experiments conducted averaged over individual datasets.}
	\label{fig:boxplot}
\end{figure*}

\section{Results and Discussion} \label{sec:results}

\subsection{Performance and comparison}

Figure \ref{fig:noise5} shows a comparison between AAI, and four baseline algorithms - SEDREAMS, DPI, MMF, ZFR on the GCI detection task for two different types of additive noise - synthetic white noise and multispeaker babble noise. We can consistently see superior performance for the AAI algorithm, across all metrics, IDR (higher is better), MR, FAR, IDA (lower is better). Both AAI and ZFR demonstrate robustness to noise even at very low SNRs. The higher performance of our algorithm may be attributed to the fact that we operate directly on the signal of interest (EGG) instead of ancillary signals such as the inverse filtered speech, or its derivatives, which degrade with noise and are severely restricted by the assumptions of the source filter model. In addition, we specifically choose the informative prior $p(\bz)$ that would encourage learning the same latent representation for both clean and noisy speech, which would help in alleviating the effects of noise. The inverse filtered approach however displays consistent deterioration with decreasing SNR (E.g., DPI algorithm). The fact that the AAI model offers the best performance despite being trained on clean speech of another dataset, vouches for its generalization capabilities.

Previous work on the task of GCI detection relies on the extraction of voiced region from the EGG signal as a preprocessing step. Our method removes this dependence on the ground truth EGG and accurately captures the voiced and unvoiced regions via a simple energy based method since it precisely estimates the voicing boundaries as shown in Fig. \ref{fig:vuv}.

Tables \ref{tab:cmu}, \ref{tab:irish}, \ref{tab:german} describe the performance of AAI on different speakers in the CMU dataset, different voice qualities in the VQ dataset and different pathology types in the Pathalogy dataset, respectively.
It can be observed that the estimated values of the quotient metrics lie within a small margin of error of the true values. Table \ref{tab:cmu} shows significant variation in CQ and SQ for different speaker characteristics, which can be used to distinguish between different speakers. Table \ref{tab:irish} shows similar variation in CQ and SQ for different voice qualities. The SQ for different pathologies vary significant as seen in \ref{tab:german}. Thus, any algorithm which achieves a good approximation to these distinguishing characteristics (such as AAI), can be utitized for the purpose for distinguishing on the basis of different speakers, voice qualities, pathology type and so on. While the proposed method is agnostic to the criteria chosen to extract different metrics, our metrics on the VQ dataset are also corroborated by the original work that created the VQ dataset \cite{liu2017comparison}. The estimated HNR for different speakers also well approximates the true HNR with a rank correlation of 1, however the HNR criterion in and of itself is not a distinguishing characteristic as the true values for all speakers are quite similar. In contrast, it has utility in distinguishing between voice qualities \cite{yumoto1982harmonics} where we observe significant difference between normal and breathy qualities and so is the case with different pathologies as well.

Figure \ref{fig:boxplot} shows the average performance of AAI across all datasets, and the gamut of defined evaluation criteria. Both GCI and GOI have nearly optimal absolute values across all datasets, with nominal decrease in median values on addition of noise. The consistent performance is a testament to the generalization of the network and the efficacy of the method proposed. There is an increase in the variation for GCI and GOI for different voice qualities and pathologies due to the inability to capture certain nuances in the EGG. Figures \ref{fig:pathology} and \ref{fig:voice_quality} affirm this claim, where AAI is unable to capture the shape at the extrema of the EGG signal. However, it is important to note that the signal characteristics are captured to a significant degree for all the pathologies and voice qualities in which the estimated and the ground truth EGG  are virtually indistinguishable unless examined closely.

Table \ref{tab:ablation} experimentally verifies our claim that the cosine loss enforces invariance to amplitude variations while retaining the temporal contours of the generated signal. The quotient metrics CQ, OQ and SQ columns clearly exhibit the inability of the norm based loss ($L_2$) to capture temporal structure accurately, where the cosine distance has much better performance. Even in tasks that depend on amplitude i.e. GCI/GOI detection, the cosine distance has performance equal to or better than the norm loss.

To evaluate the meta performance of obtaining the EGG and underscore its utility in various tasks, we demonstrate the classification performance of a shallow fully connected neural network in distinguishing between genders (on CMU) and different voice qualities (on VQ) with a train test split of 70:30  in Table \ref{tab:gender_vq_classify}. The two cases consider either glottal parameters or the EGG itself as input features, and both inputs served as excellent predictors for the tasks outlined. Every input frame is considered as an independent data point for this study. Furthermore, we also computed a pointwise distance to compare the ground truth with the predicted EGG which yielded a L2 norm (averaged across windows) of $0.0045$, indicating that the predicted EGG is in fact \textit{close} to the actual signal.

\begin{table}[]
	\centering
	\caption{Classification performance with glottal parameters (CQ, SQ) and EGG on CMU dataset.}
	\label{tab:gender_vq_classify}
	\begin{tabular}{@{}ccc@{}}
	\toprule
	\diagbox[width=1.8cm]{Label}{Feature} & Parameters & EGG \\ \midrule
	Gender & 99.59\% & 94.37\% \\
	Voice Quality & 99.31\% & 92.61\% \\ \bottomrule
	\end{tabular}
	\end{table}

\begin{figure}[h!]
	\includegraphics[width=0.45\textwidth]{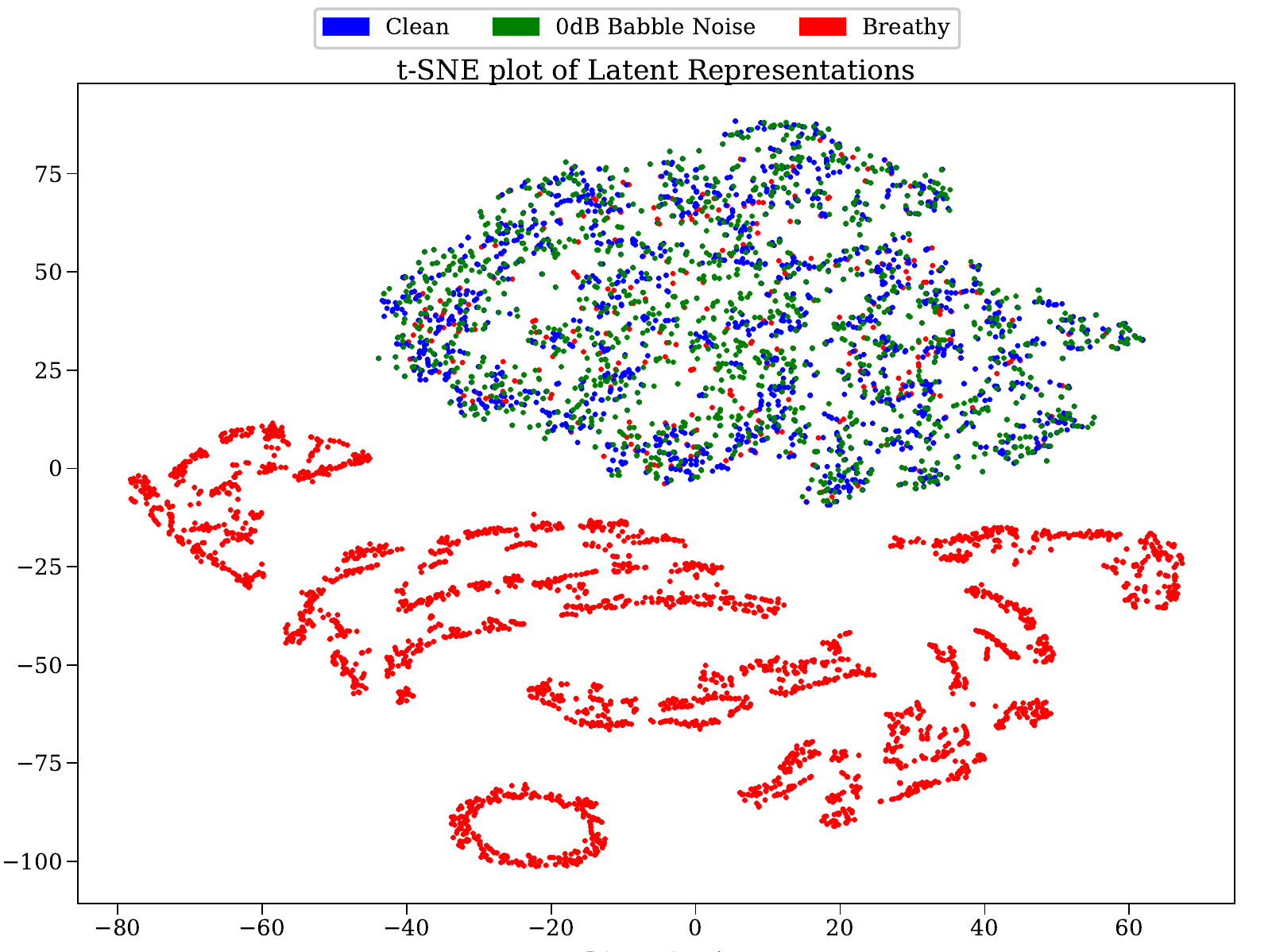}
	\centering
	\caption{Illustration of the two dimensional projections using t-SNE of the latent layer $Z$ representations for different voice qualities (CMU and VQ) and different SNRs (clean speech and corrupted with 0 dB babble noise on the CMU dataset). It can be seen that both noisy and clean speech have inseparable clusters while different voice qualities have been completely separated in the latent space.}
	\label{fig:tsne}
\end{figure}

\begin{table*}[h!]
	\centering
	\caption{Speaker-wise results on the CMU Arctic databases.}
	\label{tab:cmu}
	\resizebox{\textwidth}{!}{%
		\begin{tabular}{@{}lllllllllllllllll@{}}
			\toprule
			    & \multicolumn{4}{c}{GCI} & \multicolumn{4}{c}{GOI} & \multicolumn{2}{c}{CQ} & \multicolumn{2}{c}{OQ} & \multicolumn{2}{c}{SQ} & \multicolumn{2}{c}{HNR}                                                                               \\ \midrule
			    & IDR (\%)                & MR (\%)                 & FAR (\%)               & IDA (ms)               & IDR (\%)               & MR (\%)                 & FAR (\%) & IDA (ms) & True & Est  & True & Est. & True & Est. & True & Est. \\
			BDL & 99.49                   & 0.32                    & 0.18                   & 0.38                   & 99.03                  & 0.55                    & 0.23     & 0.51     & 0.53 & 0.53 & 0.46 & 0.46 & 1.14 & 1.05 & 2.33 & 2.42 \\
			JMK & 98.24                   & 1.39                    & 0.35                   & 0.44                   & 98.25                  & 1.55                    & 0.19     & 0.44     & 0.33 & 0.39 & 0.66 & 0.60 & 1.31 & 1.25 & 2.41 & 2.55 \\
			SLT & 99.52                   & 0.18                    & 0.29                   & 0.35                   & 99.46                  & 0.22                    & 0.31     & 0.52     & 0.40 & 0.44 & 0.59 & 0.55 & 0.76 & 0.91 & 2.39 & 2.45 \\ \bottomrule
		\end{tabular}%
	}
\end{table*}

\begin{table*}[h!]
	\centering
	\caption{Results on the VQ dataset split over three different voice qualities.}
	\label{tab:irish}
	\resizebox{\textwidth}{!}{%
		\begin{tabular}{@{}lllllllllllllllll@{}}
			\toprule
			        & \multicolumn{4}{c}{GCI} & \multicolumn{4}{c}{GOI} & \multicolumn{2}{c}{CQ} & \multicolumn{2}{c}{OQ} & \multicolumn{2}{c}{SQ} & \multicolumn{2}{c}{HNR}                                                                               \\ \midrule
			        & IDR (\%)                & MR (\%)                 & FAR (\%)               & IDA (ms)               & IDR (\%)               & MR (\%)                 & FAR (\%) & IDA (ms) & True & Est  & True & Est. & True & Est. & True & Est. \\
			Breathy & 90.94                   & 1.89                    & 7.16                   & 0.41                   & 90.30                  & 1.06                    & 8.64     & 0.30     & 0.32 & 0.33 & 0.67 & 0.66 & 1.21 & 1.24 & 2.14 & 2.37 \\
			Normal  & 92.67                   & 5.14                    & 2.19                   & 1.70                   & 92.52                  & 4.08                    & 3.39     & 0.66     & 0.44 & 0.43 & 0.55 & 0.56 & 1.84 & 1.81 & 2.81 & 2.67 \\
			Pressed & 94.40                   & 3.69                    & 1.92                   & 1.10                   & 91.79                  & 4.99                    & 3.22     & 0.70     & 0.53 & 0.51 & 0.46 & 0.48 & 1.76 & 1.87 & 2.83 & 2.51 \\ \bottomrule
		\end{tabular}%
	}
\end{table*}

\begin{table*}[h!]
	\centering
	\caption{Results on eight different pathologies from the Pathology dataset.}
	\label{tab:german}
	\resizebox{\textwidth}{!}{%
		\begin{tabular}{@{}lllllllllllllllll@{}}
			\toprule
			                              & \multicolumn{4}{c}{GCI} & \multicolumn{4}{c}{GOI} & \multicolumn{2}{c}{CQ} & \multicolumn{2}{c}{OQ} & \multicolumn{2}{c}{SQ} & \multicolumn{2}{c}{HNR}                                                                               \\ \midrule
			                              & IDR (\%)                & MR (\%)                 & FAR (\%)               & IDA (ms)               & IDR (\%)               & MR (\%)                 & FAR (\%) & IDA (ms) & True & Est  & True & Est. & True & Est. & True & Est. \\
			Balbuties                     & 98.54                   & 0.74                    & 0.72                   & 0.64                   & 97.83                  & 1.08                    & 1.08     & 0.27     & 0.45 & 0.42 & 0.55 & 0.58 & 2.25 & 2.02 & 2.30 & 2.22 \\
			Diplophonie                   & 92.94                   & 2.73                    & 4.33                   & 0.98                   & 92.83                  & 3.06                    & 4.11     & 0.88     & 0.42 & 0.39 & 0.58 & 0.61 & 1.72 & 1.76 & 2.53 & 2.66 \\
			Dysodie                       & 94.00                   & 2.38                    & 3.66                   & 1.30                   & 89.88                  & 4.48                    & 5.64     & 1.13     & 0.43 & 0.42 & 0.57 & 0.58 & 2.00 & 2.09 & 1.88 & 2.01 \\
			GERD                          & 99.33                   & 0.00                    & 0.67                   & 0.69                   & 98.69                  & 0.00                    & 1.31     & 0.19     & 0.43 & 0.41 & 0.57 & 0.59 & 2.34 & 2.49 & 2.53 & 2.86 \\
			Amyotrophic Lateral Sclerosis & 100.00                  & 0.00                    & 0.00                   & 0.55                   & 99.55                  & 0.45                    & 0.00     & 0.26     & 0.40 & 0.36 & 0.60 & 0.64 & 1.32 & 1.45 & 1.90 & 1.70 \\
			Aryluxation                   & 91.80                   & 2.31                    & 5.89                   & 1.65                   & 84.85                  & 7.54                    & 7.61     & 1.51     & 0.43 & 0.43 & 0.57 & 0.57 & 1.96 & 2.15 & 2.34 & 2.62 \\
			Intubation Damage             & 93.55                   & 3.21                    & 3.24                   & 1.24                   & 85.06                  & 7.45                    & 7.48     & 1.10     & 0.43 & 0.39 & 0.57 & 0.61 & 1.56 & 1.67 & 1.48 & 2.67 \\
			Gesangsstimme                 & 91.06                   & 4.46                    & 4.48                   & 1.48                   & 85.23                  & 7.36                    & 7.41     & 1.35     & 0.42 & 0.44 & 0.58 & 0.56 & 2.24 & 2.36 & 2.43 & 2.37 \\
			\bottomrule
		\end{tabular}%
	}
\end{table*}

\begin{table*}[h!]
	\centering
	\caption{Comparison of the two approximations for the likelihood loss term on CMU Arctic datasets.}
	\label{tab:ablation}
	\resizebox{\textwidth}{!}{%
		\begin{tabular}{@{}lllllllllllllllll@{}}
			\toprule
			             & \multicolumn{4}{c}{GCI} & \multicolumn{4}{c}{GOI} & \multicolumn{2}{c}{CQ} & \multicolumn{2}{c}{OQ} & \multicolumn{2}{c}{SQ} & \multicolumn{2}{c}{HNR}                                                                               \\ \midrule
			             & IDR (\%)                & MR (\%)                 & FAR (\%)               & IDA (ms)               & IDR (\%)               & MR (\%)                 & FAR (\%) & IDA (ms) & True & Est  & True & Est. & True & Est. & True & Est. \\
			Cosine dist. & 99.07                   & 0.65                    & 0.28                   & 0.39                   & 98.98                  & 0.77                    & 0.25     & 0.5      & 0.56 & 0.54 & 0.44 & 0.46 & 1.27 & 1.07 & 2.01 & 2.33 \\
			L2 dist.     & 94.90                   & 2.71                    & 2.39                   & 0.38                   & 94.81                  & 2.79                    & 2.39     & 0.73     & 0.58 & 0.52 & 0.42 & 0.48 & 1.07 & 1.65 & 2.01 & 1.63 \\ \bottomrule
		\end{tabular}%
	}
\end{table*}

\subsection{Comparison with Glottal parameter estimators}

To further test the efficacy of the proposed method, in this section, we compare it with three schemes for glottal inverse filtering, Iterative Adaptive Inverse Filtering (IAIF) \cite{alku1992glottal}, Probabilistic Weighted Linear Prediction (PWLP) \cite{rao2019glottal} and Quasi Closed Phase method (QCP) \cite{airaksinen2014quasi}, Conditional Variational Autoencoders \cite{sohn2015learning} (CVAE) and an ordinary multi-layer perceptron regressing on the parameters CQ, SQ and HNR, called Parameter MLP (PMLP). While the inverse filtering based techniques estimate the glottal flow, CVAE and PMLP estimate the EGG and the glottal parameters, respectively. Both QCP and PWLP use weighted linear prediction which is considered more robust to the harmonic structure of speech as compared to standard linear prediction analysis (LP). IAIF uses a multi step iterative sequence of filters to estimate glottal flow and vocal tract envelopes, whose primary advantage is the lack of requirement of any GCI/GOI information. Both QCP and PWLP require GCI/GOI information, however PWLP estimates the closed phase of the source directly from speech instead of GCI location, allowing for better estimation of the voice source.
	On the other hand in a CVAE \cite{sohn2015learning}, a conventional variational autoendoer is built on the EGG signal with an additional conditioning of the speech signal in the latent space that is forced to be a Normal distribution. During inference, only the Decoder is used by conditioning it with the input speech segment to obtain the corresponding EGG signal at the output. The key difference between AAI and a CVAE is that we impose an informative prior (derived from the EGG space) on the latent space through KL minimization achieved via adversarial learning while in a CVAE, the latent space is Normally distributed. Further, since in our model, the the aggregated prior is matched (and not the conditional prior unlike the CVAE), the problems associated with the VAE in simultaneously maximizing the likelihood and conditional KL minimization \cite{dai2019diagnosing} do not exist. These changes, we believe will lead to better generalization as demonstrated through empirical evidence in the subsequent paragraphs). For AAI, CVAE and PMLP, the Childers dataset has been used for training and CMU (with and without noise) and VQ datasets for testing.


\begin{table}[h!]
	\centering{
		\caption{Comparison of AAI with several glottal parameter estimation techniques on CMU dataset (clean).}
		\label{tab:met_inv_cmu_clean}
		\resizebox{0.45\textwidth}{!}{%
			\begin{tabular}{@{}cccccccc@{}}
				\toprule
				    & GND           & AAI           & IAIF          & PWLP & QCP  & CVAE & PMLP \\ \midrule
				CQ  & \textit{0.42} & \textbf{0.45} & \textbf{0.45} & 0.51 & 0.52 & 0.46 & 0.28 \\
				SQ  & \textit{1.07} & \textbf{1.07} & 1.35          & 1.22 & 1.12 & 0.85 & 1.78 \\
				HNR & \textit{2.37} & \textbf{2.47} & 1.92          & 1.41 & 1.31 & 2.22 & 2.49 \\ \bottomrule
			\end{tabular}%
		}}
\end{table}

\begin{table}[]
	\centering{
		\caption{Comparison of AAI with several techniques on CMU dataset (with 0 dB babble noise).}
		\label{tab:cmu_babble_inv_glot}
		\resizebox{0.45\textwidth}{!}{%
			\begin{tabular}{@{}cccccccc@{}}
				\toprule
				    & GND           & AAI           & IAIF & PWLP & QCP  & CVAE & PMLP          \\ \midrule
				CQ  & \textit{0.42} & \textbf{0.44} & 0.46 & 0.52 & 0.52 & 0.52 & 0.28          \\
				SQ  & \textit{1.07} & \textbf{1.02} & 1.54 & 1.68 & 1.48 & 1.15 & 1.82          \\
				HNR & \textit{2.37} & \textbf{2.49} & 1.78 & 1.87 & 1.49 & 1.71 & \textbf{2.55} \\ \bottomrule
			\end{tabular}%
		}}
\end{table}

\begin{table}[h!]
	\centering{
		\caption{Comparison of AAI with several glottal parameter estimators on CMU dataset (with 0 dB White noise)}
		\label{tab:cmu_white_inv_glot}
		\resizebox{0.45\textwidth}{!}{%
			\begin{tabular}{@{}cccccccc@{}}
				\toprule
				    & GND            & AAI           & IAIF          & PWLP & QCP  & CVAE & PMLP \\ \midrule
				CQ  & \textit{ 0.42} & \textbf{0.46} & \textbf{0.46} & 0.52 & 0.52 & 0.52 & 0.28 \\
				SQ  & \textit{1.07}  & \textbf{0.94} & 1.55          & 1.35 & 1.46 & 1.23 & 1.83 \\
				HNR & \textit{2.37}  & \textbf{2.45} & 1.77          & 1.58 & 1.26 & 1.75 & 2.52 \\ \bottomrule
			\end{tabular}%
		}}
\end{table}

\begin{table}[]
	\centering{
		\caption{Comparison of AAI with several glottal flow estimation techniques on the Voice Quality dataset}
		\label{tab:irish_inv_glot}
		\resizebox{0.45\textwidth}{!}{%
			\begin{tabular}{@{}cccccccc@{}}
				\toprule
				    & GND           & AAI           & IAIF & PWLP & QCP  & CVAE & PMLP \\ \midrule
				CQ  & \textit{0.56} & \textbf{0.57} & 0.47 & 0.48 & 0.49 & 0.53 & 0.23 \\
				SQ  & \textit{1.60} & \textbf{1.64} & 2.82 & 4.02 & 3.61 & 1.22 & 3.82 \\
				HNR & \textit{2.75} & \textbf{2.77} & 1.41 & 3.87 & 4.49 & 2.52 & 1.34 \\ \bottomrule
			\end{tabular}%
		}}
\end{table}


Tables \ref{tab:met_inv_cmu_clean}, \ref{tab:cmu_babble_inv_glot}, \ref{tab:cmu_white_inv_glot} compare the performance of AAI against baseline schemes on the CMU dataset, for clean speech and speech corrupted by 0 dB babble and 0 dB white noise respectively.

The inverse filtering techniques use highly restrictive model assumptions on the voicing process such as assuming the voice filter to be an all pole LTI system. The learning based methods such as CVAE and PMLP relax these model assumptions, however even they are susceptible to the presence of noise in the input signal, as unlike AAI, they do not enforce learning representations invariant to input noise.

It can be seen that AAI matches or outperforms baseline schemes on all cases. All inverse filtering methods have deteriorating performance at higher formants which is reflected in the worse SQ and HNR values.
	AAI also outperforms the deep learning methods, CVAE and PMLP. PMLP fails to generalise due to the different representations learnt for different kinds of noise corrupted speech, while AAI enforces learning noise invariant representations for speech. This leads to consistently worse performance in CVAE and PMLP parameter estimation when inferring on speech characteristics different from the training set.
	Table \ref{tab:irish_inv_glot} corroborates this, where the network PMLP trained on Childers data fails to generalise to the VQ dataset, where the speech characteristics are substantially different from the Childer's data. The consistently superior empirical performance of AAI in parameter estimation across all datasets, further substantiates that the EGG signal is an optimal representation for such tasks as opposed to voice source estimation techniques.

\subsection{Analysis of Latent Representation}

Figure \ref{fig:tsne} demonstrates the crux of Adversarial Approximate Inference and its effect on the learnt latent representation. The scatter plot in Fig. \ref{fig:tsne} represents the projection of the learnt latent vectors onto a 2-dimensional plane computed using t-SNE \cite{maaten2008visualizing}. The first feature to be noted is the distinct clusters formed by the different voice qualities in the VQ Dataset and the CMU dataset, which implies that the distribution over the embeddings of the VQ speech samples $\mathbb{P}(\bz \mid \text{VQ speech})$ and  CMU speech samples $\mathbb{P}(\bz \mid \text{CMU speech})$ have different support in the latent space and the network successfully disentangles the factors of variation in the two kinds of voices.

Secondly, the AAI framework motivated by the need for informative priors and robustness to noise in the input signal, enforces learning a representation that discards information that is irrelevant to produce the output. This is demonstrated by the embeddings of clean and noisy speech for the CMU Dataset in Fig. \ref{fig:tsne} where both clean and noisy are inseparable in the latent space and the projections. It is desirous to learn a similar latent representation for both noisy and clean speech, since the output of the model i.e. the laryngograph signal remains the same in both scenarios. Both these features aptly demonstrate the efficacy of our technique in ameliorating the effects of noise by discarding noise in the input signal, while successfully retaining information that can distinguish between different auditory colorings (or voice qualities).

\begin{figure}[h!]
	\includegraphics[width=0.45\textwidth, height = 1.7 in]{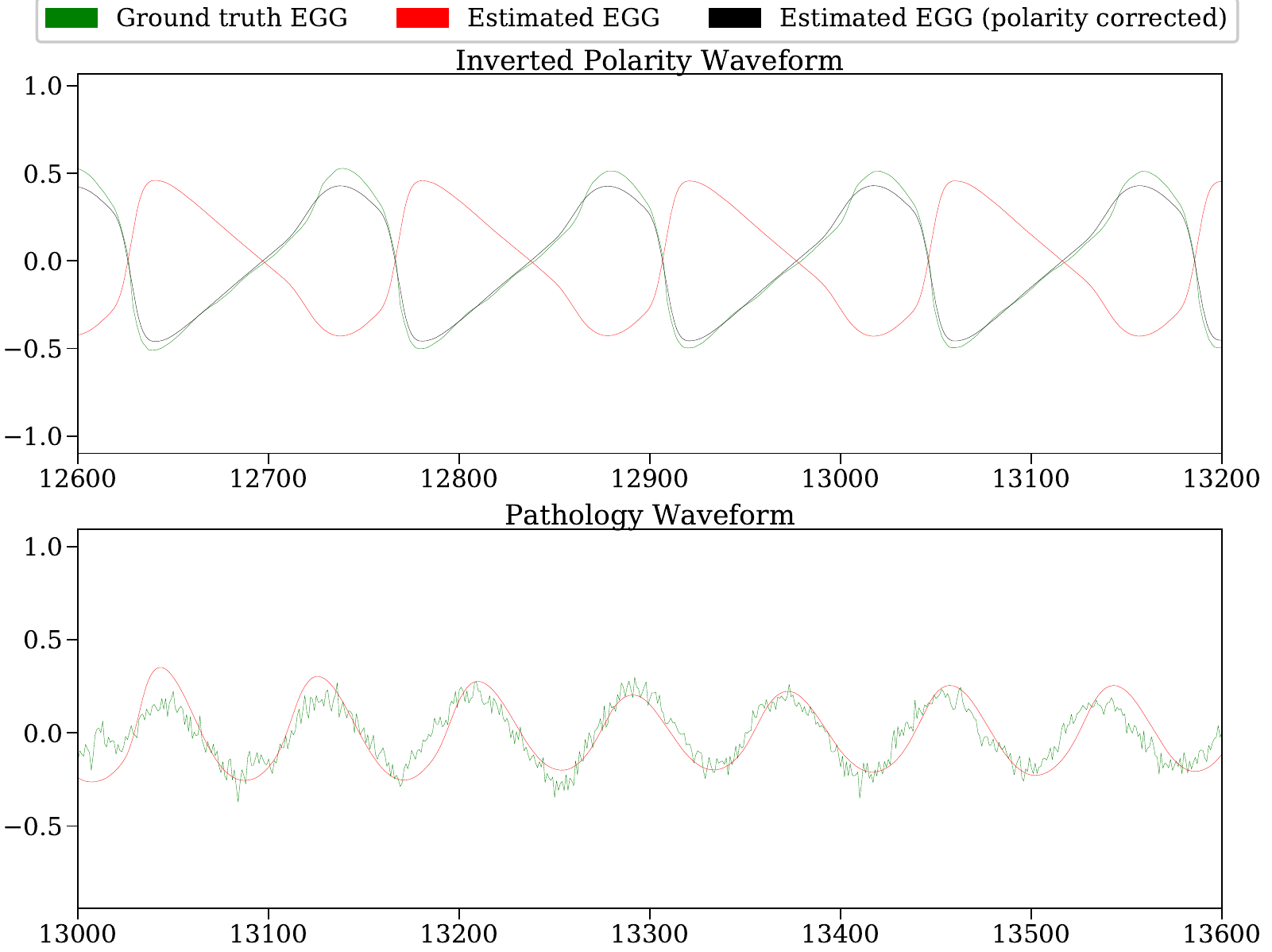}
	\centering
	\caption{Illustration of shortcomings of the AAI method on inverted speech polarity and the pathological case with significant high-frequency noise.}
	\label{fig:limitation}
\end{figure}

\subsection{Limitations}
In order to evaluate its robustness, AAI's efficacy was demonstrated across a number of tasks and across the voice diaspora. However, being a non linear processing system, the method is still susceptible to the problem of incorrect speech polarity. The asymmetry of the glottal waveform implies that the speech signal is also asymmetric \cite{drugman2013residual}, which manifests as different speech polarities when measuring via a microphone. Figure \ref{fig:limitation} demonstrates this issue, where we have inverted polarities for speech and the estimated EGG, as our method assumes a positive polarity for the speech signal, defined in \cite{drugman2013residual}. Where applicable, we have manually corrected for such artifacts by inverting the speech polarity. Part (b) in Fig. \ref{fig:limitation} displays a segment of a pathology where AAI fails to capture the higher order harmonics. For robustness to noise, our method utilizes singly strided frames across the speech signal, which consequently acts as a smoothing operation and severely attenuates higher order harmonics. This can be a potential limitation when working with voices which inherently have significant amplitude at high harmonics. However, the method can be made to capture such signal behaviours by having non-overlapping windows.

\subsection{Conclusion} \label{sec:conclusion}

We proposed a distribution transformation framework to map speech to the corresponding EGG signal, that is robust to noise, and generalizes across recording conditions, speech pathologies and voice qualities.
In essence, AAI is a unifying framework for the complete class of methods that create task specific representations and techniques for exploiting the information available in the electroglottographic signal.
While the efficacy of AAI is empirically verified in the setting of speech transformation, the constructed framework is agnostic to the application chosen, and can be used in an array of problems. Since learning conditional distributions is a task of ubiquitous importance, future work on this framework can focus on other areas in which similar principles may be applied, and further investigate the statistical properties of the latent space constructed by our model.

\section*{Acknowledgment}

The authors would like to thank Dr. T V Ananthapadmanabha for his vaulable comments on the manuscript, Prof. Prasanta Kumar Ghosh for lending the Childer's dataset, Prof. Anne-Maria Laukkanen for sharing the VQ dataset. They would also like to thank the authors of IAIF, QCP, PWLP and CVAE for providing the codes for their implementations. They would thank the associate editor and the reviewers who helped to enhance the quality of the work significantly.

\ifCLASSOPTIONcaptionsoff
	\newpage
\fi



\bibliographystyle{ieeetr}
\bibliography{references}

\begin{thebibliography}{10}

\bibitem{laver1994principles}
J.~Laver and L.~John, {\em Principles of phonetics}.
\newblock Cambridge University Press, 1994.

\bibitem{marasek1997egg}
K.~Marasek, ``Egg and voice quality,'' {\em Universit{\"a}t Stuttgart:
  http://www. ims. uni-stuttgart. de/phonetik/EGG/frmst1. htm}, 1997.

\bibitem{trask2004dictionary}
R.~L. Trask, {\em A dictionary of phonetics and phonology}.
\newblock Routledge, 2004.

\bibitem{ladefoged1988investigating}
P.~Ladefoged, I.~Maddieson, and M.~Jackson, ``Investigating phonation types in
  different languages,'' {\em Vocal physiology: Voice production, mechanisms
  and functions}, pp.~297--317, 1988.

\bibitem{hirano1981clinical}
M.~Hirano, ``Clinical examination of voice,'' {\em Disorders of human
  communication}, vol.~5, pp.~1--99, 1981.

\bibitem{kitzing1985stroboscopy}
P.~Kitzing, ``Stroboscopy--a pertinent laryngological examination.,'' {\em The
  Journal of Otolaryngology}, vol.~14, no.~3, pp.~151--157, 1985.

\bibitem{gobl1992acoustic}
C.~Gobl and A.~N. Chasaide, ``Acoustic characteristics of voice quality,'' {\em
  Speech Communication}, vol.~11, no.~4-5, pp.~481--490, 1992.

\bibitem{wong1979least}
D.~Wong, J.~Markel, and A.~Gray, ``Least squares glottal inverse filtering from
  the acoustic speech waveform,'' {\em IEEE Transactions on Acoustics, Speech,
  and Signal Processing}, vol.~27, no.~4, pp.~350--355, 1979.

\bibitem{wakita1973direct}
H.~Wakita, ``Direct estimation of the vocal tract shape by inverse filtering of
  acoustic speech waveforms,'' {\em IEEE Transactions on Audio and
  Electroacoustics}, vol.~21, no.~5, pp.~417--427, 1973.

\bibitem{airaksinen2014quasi}
M.~Airaksinen, T.~Raitio, B.~Story, and P.~Alku, ``Quasi closed phase glottal
  inverse filtering analysis with weighted linear prediction,'' {\em IEEE/ACM
  Transactions on Audio, Speech and Language Processing (TASLP)}, vol.~22,
  no.~3, pp.~596--607, 2014.

\bibitem{airaksinen2017quadratic}
M.~Airaksinen, T.~B{\"a}ckstr{\"o}m, and P.~Alku, ``Quadratic programming
  approach to glottal inverse filtering by joint norm-1 and norm-2
  optimization,'' {\em IEEE/ACM Transactions on Audio, Speech, and Language
  Processing}, vol.~25, no.~5, pp.~929--939, 2017.

\bibitem{rao2019glottal}
A.~Rao and P.~K. Ghosh, ``Glottal inverse filtering using probabilistic
  weighted linear prediction,'' {\em IEEE/ACM Transactions on Audio, Speech,
  and Language Processing}, vol.~27, no.~1, pp.~114--124, 2019.

\bibitem{raitio2011hmm}
T.~Raitio, A.~Suni, J.~Yamagishi, H.~Pulakka, J.~Nurminen, M.~Vainio, and
  P.~Alku, ``Hmm-based speech synthesis utilizing glottal inverse filtering,''
  {\em IEEE Transactions on Audio, Speech, and Language Processing}, vol.~19,
  no.~1, pp.~153--165, 2011.

\bibitem{makhoul1975linear}
J.~Makhoul, ``Linear prediction: A tutorial review,'' {\em Proceedings of the
  IEEE}, vol.~63, no.~4, pp.~561--580, 1975.

\bibitem{alku2011glottal}
P.~Alku, ``Glottal inverse filtering analysis of human voice production—a
  review of estimation and parameterization methods of the glottal excitation
  and their applications,'' {\em Sadhana}, vol.~36, no.~5, pp.~623--650, 2011.

\bibitem{veeneman1985automatic}
D.~Veeneman and S.~BeMent, ``Automatic glottal inverse filtering from speech
  and electroglottographic signals,'' {\em IEEE transactions on acoustics,
  speech, and signal processing}, vol.~33, no.~2, pp.~369--377, 1985.

\bibitem{baken1992electroglottography}
R.~J. Baken, ``Electroglottography,'' {\em Journal of Voice}, vol.~6, no.~2,
  pp.~98--110, 1992.

\bibitem{childers1985critical}
D.~G. Childers and A.~K. Krishnamurthy, ``A critical review of
  electroglottography.,'' {\em Critical reviews in biomedical engineering},
  vol.~12, no.~2, pp.~131--161, 1985.

\bibitem{titze1990interpretation}
I.~R. Titze, ``Interpretation of the electroglottographic signal,'' {\em
  Journal of Voice}, vol.~4, no.~1, pp.~1--9, 1990.

\bibitem{peterson1994comparison}
K.~L. Peterson, K.~Verdolini-Marston, J.~M. Barkmeier, and H.~T. Hoffman,
  ``Comparison of aerodynamic and electroglottographic parameters in evaluating
  clinically relevant voicing patterns,'' {\em Annals of Otology, Rhinology \&
  Laryngology}, vol.~103, no.~5, pp.~335--346, 1994.

\bibitem{liu2017comparison}
D.~Liu, E.~Kankare, A.-M. Laukkanen, and P.~Alku, ``Comparison of
  parametrization methods of electroglottographic and inverse filtered acoustic
  speech pressure signals in distinguishing between phonation types,'' {\em
  Biomedical Signal Processing and Control}, vol.~36, pp.~183--193, 2017.

\bibitem{chen2002electroglottographic}
Y.~Chen, M.~P. Robb, and H.~R. Gilbert, ``Electroglottographic evaluation of
  gender and vowel effects during modal and vocal fry phonation,'' {\em Journal
  of speech, language, and hearing research}, 2002.

\bibitem{higgins2002gender}
M.~B. Higgins and L.~Schulte, ``Gender differences in vocal fold contact
  computed from electroglottographic signals: the influence of measurement
  criteria,'' {\em The Journal of the Acoustical Society of America}, vol.~111,
  no.~4, pp.~1865--1871, 2002.

\bibitem{waaramaa2013acoustic}
T.~Waaramaa and E.~Kankare, ``Acoustic and egg analyses of emotional
  utterances,'' {\em Logopedics Phoniatrics Vocology}, vol.~38, no.~1,
  pp.~11--18, 2013.

\bibitem{murphy2009electroglottogram}
P.~J. Murphy and A.-M. Laukkanen, ``Electroglottogram analysis of emotionally
  styled phonation,'' in {\em Multimodal signals: Cognitive and algorithmic
  issues}, pp.~264--270, Springer, 2009.

\bibitem{schered1987electroglottography}
R.~SCHERED, ``Electroglottography and direct measurement of vocal fold contact
  area,'' {\em Vocal Physiology Voice production mechanisms and functions},
  pp.~279--291, 1987.

\bibitem{deshpande2018effective}
P.~S. Deshpande and M.~S. Manikandan, ``Effective glottal instant detection and
  electroglottographic parameter extraction for automated voice pathology
  assessment,'' {\em IEEE journal of biomedical and health informatics},
  vol.~22, no.~2, pp.~398--408, 2018.

\bibitem{kitzing1990clinical}
P.~Kitzing, ``Clinical applications of electroglottography,'' {\em Journal of
  Voice}, vol.~4, no.~3, pp.~238--249, 1990.

\bibitem{alku2003parameterisation}
P.~Alku, ``Parameterisation methods of the glottal flow estimated by inverse
  filtering,'' in {\em ISCA Tutorial and Research Workshop on Voice Quality:
  Functions, Analysis and Synthesis}, 2003.

\bibitem{drugman2012comparative}
T.~Drugman, B.~Bozkurt, and T.~Dutoit, ``A comparative study of glottal source
  estimation techniques,'' {\em Computer Speech \& Language}, vol.~26, no.~1,
  pp.~20--34, 2012.

\bibitem{murty2008epoch}
K.~S.~R. Murty and B.~Yegnanarayana, ``Epoch extraction from speech signals,''
  {\em IEEE Transactions on Audio, Speech, and Language Processing}, vol.~16,
  no.~8, pp.~1602--1613, 2008.

\bibitem{drugman2008voice}
T.~Drugman, T.~Dubuisson, N.~D'Alessandro, A.~Moinet, and T.~Dutoit, ``Voice
  source parameters estimation by fitting the glottal formant and the inverse
  filtering open phase,'' in {\em Signal Processing Conference, 2008 16th
  European}, pp.~1--5, IEEE, 2008.

\bibitem{prathosh2013epoch}
A.~Prathosh, T.~Ananthapadmanabha, and A.~Ramakrishnan, ``Epoch extraction
  based on integrated linear prediction residual using plosion index,'' {\em
  IEEE Transactions on Audio, Speech, and Language Processing}, vol.~21,
  no.~12, pp.~2471--2480, 2013.

\bibitem{drugman2012detection}
T.~Drugman, M.~Thomas, J.~Gudnason, P.~Naylor, and T.~Dutoit, ``Detection of
  glottal closure instants from speech signals: A quantitative review,'' {\em
  IEEE Transactions on Audio, Speech, and Language Processing}, vol.~20, no.~3,
  pp.~994--1006, 2012.

\bibitem{alku1992glottal}
P.~Alku, ``Glottal wave analysis with pitch synchronous iterative adaptive
  inverse filtering,'' {\em Speech communication}, vol.~11, no.~2-3,
  pp.~109--118, 1992.

\bibitem{goodfellow2014generative}
I.~Goodfellow, J.~Pouget-Abadie, M.~Mirza, B.~Xu, D.~Warde-Farley, S.~Ozair,
  A.~Courville, and Y.~Bengio, ``Generative adversarial nets,'' in {\em
  Advances in neural information processing systems}, pp.~2672--2680, 2014.

\bibitem{goodfellow2016nips}
I.~Goodfellow, ``Nips 2016 tutorial: Generative adversarial networks,'' {\em
  arXiv preprint arXiv:1701.00160}, 2016.

\bibitem{kingma2013auto}
D.~P. Kingma and M.~Welling, ``Auto-encoding variational bayes,'' {\em arXiv
  preprint arXiv:1312.6114}, 2013.

\bibitem{chen2016variational}
X.~Chen, D.~P. Kingma, T.~Salimans, Y.~Duan, P.~Dhariwal, J.~Schulman,
  I.~Sutskever, and P.~Abbeel, ``Variational lossy autoencoder,'' {\em arXiv
  preprint arXiv:1611.02731}, 2016.

\bibitem{mirza2014conditional}
M.~Mirza and S.~Osindero, ``Conditional generative adversarial nets,'' {\em
  arXiv preprint arXiv:1411.1784}, 2014.

\bibitem{gulrajani2017improved}
I.~Gulrajani, F.~Ahmed, M.~Arjovsky, V.~Dumoulin, and A.~C. Courville,
  ``Improved training of wasserstein gans,'' in {\em Advances in Neural
  Information Processing Systems}, pp.~5767--5777, 2017.

\bibitem{burgess2018understanding}
C.~P. Burgess, I.~Higgins, A.~Pal, L.~Matthey, N.~Watters, G.~Desjardins, and
  A.~Lerchner, ``Understanding disentangling in $beta$-vae,'' {\em arXiv
  preprint arXiv:1804.03599}, 2018.

\bibitem{kingma2014semi}
D.~P. Kingma, S.~Mohamed, D.~J. Rezende, and M.~Welling, ``Semi-supervised
  learning with deep generative models,'' in {\em Advances in neural
  information processing systems}, pp.~3581--3589, 2014.

\bibitem{rezende2014stochastic}
D.~J. Rezende, S.~Mohamed, and D.~Wierstra, ``Stochastic backpropagation and
  approximate inference in deep generative models,'' {\em arXiv preprint
  arXiv:1401.4082}, 2014.

\bibitem{makhzani2015adversarial}
A.~Makhzani, J.~Shlens, N.~Jaitly, I.~Goodfellow, and B.~Frey, ``Adversarial
  autoencoders,'' {\em arXiv preprint arXiv:1511.05644}, 2015.

\bibitem{childers2000speech}
D.~G. Childers and J.~A. Diaz, ``Speech processing and synthesis toolboxes,''
  2000.

\bibitem{kominek2004cmu}
J.~Kominek and A.~W. Black, ``The cmu arctic speech databases,'' in {\em Fifth
  ISCA workshop on speech synthesis}, 2004.

\bibitem{woldert2007saarbruecken}
B.~Woldert-Jokisz, ``Saarbruecken voice database,'' 2007.

\bibitem{ananthapadmanabha1979epoch}
T.~Ananthapadmanabha and B.~Yegnanarayana, ``Epoch extraction from linear
  prediction residual for identification of closed glottis interval,'' {\em
  IEEE transactions on acoustics speech and signal processing}, vol.~27, no.~4,
  pp.~309--319, 1979.

\bibitem{hillman1989objective}
R.~E. Hillman, E.~B. Holmberg, J.~S. Perkell, M.~Walsh, and C.~Vaughan,
  ``Objective assessment of vocal hyperfunction: An experimental framework and
  initial results,'' {\em Journal of Speech, Language, and Hearing Research},
  vol.~32, no.~2, pp.~373--392, 1989.

\bibitem{teixeira2013vocal}
J.~P. Teixeira, C.~Oliveira, and C.~Lopes, ``Vocal acoustic analysis--jitter,
  shimmer and hnr parameters,'' {\em Procedia Technology}, vol.~9,
  pp.~1112--1122, 2013.

\bibitem{khanagha2014detection}
V.~Khanagha, K.~Daoudi, and H.~M. Yahia, ``Detection of glottal closure
  instants based on the microcanonical multiscale formalism,'' {\em IEEE/ACM
  Transactions on Audio, Speech and Language Processing (TASLP)}, vol.~22,
  no.~12, pp.~1941--1950, 2014.

\bibitem{drugman2009glottal}
T.~Drugman and T.~Dutoit, ``Glottal closure and opening instant detection from
  speech signals,'' in {\em Tenth Annual Conference of the International Speech
  Communication Association}, 2009.

\bibitem{yumoto1982harmonics}
E.~Yumoto, W.~J. Gould, and T.~Baer, ``Harmonics-to-noise ratio as an index of
  the degree of hoarseness,'' {\em The journal of the Acoustical Society of
  America}, vol.~71, no.~6, pp.~1544--1550, 1982.

\bibitem{sohn2015learning}
K.~Sohn, H.~Lee, and X.~Yan, ``Learning structured output representation using
  deep conditional generative models,'' in {\em Advances in Neural Information
  Processing Systems}, pp.~3483--3491, 2015.

\bibitem{dai2019diagnosing}
B.~Dai and D.~Wipf, ``Diagnosing and enhancing vae models,'' {\em arXiv
  preprint arXiv:1903.05789}, 2019.

\bibitem{maaten2008visualizing}
L.~v.~d. Maaten and G.~Hinton, ``Visualizing data using t-sne,'' {\em Journal
  of machine learning research}, vol.~9, no.~Nov, pp.~2579--2605, 2008.

\bibitem{drugman2013residual}
T.~Drugman, ``Residual excitation skewness for automatic speech polarity
  detection,'' {\em IEEE Signal Processing Letters}, vol.~20, no.~4,
  pp.~387--390, 2013.

\end{thebibliography}








\end{document}